\documentclass[12pt]{article}
\usepackage[letterpaper,top=2cm,bottom=2cm,left=2cm,right=2cm,marginparwidth=2cm]{geometry}
\usepackage{graphicx} 
\usepackage{amsmath, amssymb}
\usepackage{amsthm}
\usepackage{siunitx}
\usepackage[font=small,margin=1.5cm]{caption}
\usepackage[dvipsnames]{xcolor}
\definecolor{2e308d}{RGB}{46,48,141}
\usepackage[colorlinks,citecolor=2e308d,filecolor=2e308d,urlcolor=2e308d,bookmarks=false,hypertexnames=true]{hyperref}
\usepackage{authblk}
\usepackage{orcidlink}
\usepackage{cite}
\usepackage{booktabs}
\usepackage{threeparttable}
\usepackage{float}
\usepackage{algorithm}
\usepackage{algorithmic}
\usepackage{braket}
\usepackage{qcircuit}
\usepackage{titlesec}

\newtheorem{lemma}{Lemma}
\newtheorem{remark}{Remark}

\definecolor{daxcolor}{rgb}{0.2, 0.1, 1}

\usepackage[T1]{fontenc}

\title{Quantum Circuits for Variable-Coefficient Vlasov Dynamics via Schr\"odingerisation}

\author[1]{Fong Yew Leong\,\orcidlink{0000-0002-0064-0118}\thanks{Corresponding author. Email: leongfy@a-star.edu.sg}} 
\author[2,3,4]{Dax Enshan Koh\,\orcidlink{0000-0002-8968-591X}} 
\author[1]{Wei-Bin Ewe\,\orcidlink{0000-0002-4600-0634}} 
\author[1]{Qiubo Chen\,\orcidlink{0000-0002-0308-8111}} 
\author[1]{Zhun-Yong Ong\,\orcidlink{0000-0003-2668-6453}} 

\affil[1]{\small \href{https://ror.org/01eczt548}{Institute of Advanced Intelligence and Computing (IAIC)}, \href{https://ror.org/036wvzt09}{Agency for Science, Technology and Research (A*STAR)}, 1 Fusionopolis Way, \#16-16 Connexis, Singapore 138632, Singapore}
\affil[2]{\small Engineering Cluster, \href{https://ror.org/01v2c2791}{Singapore Institute of Technology}, 1 Punggol Coast Road, Singapore 828608, Singapore}
\affil[3]{\small
Quantum Innovation Centre (Q.InC), \href{https://ror.org/036wvzt09}{Agency for Science, Technology and Research (A*STAR)}, 4 Fusionopolis Way, Kinesis \#05-01, Singapore 138635, Republic of Singapore}
\affil[4]{\small Science, Mathematics and Technology Cluster, \href{https://ror.org/05j6fvn87}{Singapore University of Technology and Design}, 8 Somapah Road, Singapore 487372, Singapore}

\date{}

\begin{document}

\maketitle

\begin{abstract}
High-dimensional Vlasov equations present promising use cases for quantum simulation, but their transport coefficients must be implemented coherently in order to preserve a compact quantum representation. We develop explicit Schr\"odingerised circuits for prescribed external field Vlasov dynamics without classical field updates. In 1D1V, velocity-dependent streaming is compiled directly from the encoded velocity register, while a non-linear cubic field is evaluated using reversible fixed-point arithmetic. For fixed polynomial degree, the field workspace scales linearly with the coordinate register width, while the field evaluation subroutine has polynomial depth. As a proof of concept, we simulate a minimal coupled 3D3V anharmonic Penning system and realise mixed electric-field terms and transverse magnetic coupling on a six-dimensional phase space in agreement with classical calculations. Resource analysis shows approximately linear growth with the number of time steps, while identifying the auxiliary register as the dominant depth bottleneck. Structured preparation and register recovery of the auxiliary state reveal an interplay between loading cost and readout. Together, our algorithm demonstrates a fully coherent quantum simulation of high-dimensional Vlasov dynamics with variable coefficients and identifies the practical resource costs that must be controlled for large-scale quantum implementation.
\end{abstract}

\vspace{2.0em}
\noindent\textbf{Keywords:} quantum simulation; Schr\"odingerisation; Vlasov equation; variable-coefficient quantum circuits; matrix product states

\section{Introduction}
\label{sec:introduction}

The Vlasov equation provides a kinetic description of collisionless plasmas, evolving a distribution function over both configuration and velocity space. In the physically standard three-dimensional setting, the distribution \(f=f(\mathbf{x},\mathbf{v},t)\) therefore occupies a six-dimensional phase space. The resulting curse of dimensionality is a central computational challenge of grid-based Vlasov simulation~\cite{Palmroth2025}. For a direct tensor-product discretisation with \(N\) points along each coordinate, the phase-space representation contains \(N^6\) values, causing memory and computational requirements to grow rapidly with resolution. Consequently, fully kinetic 4D--6D Vlasov calculations have required increasingly large high-performance computing resources, including implementations on large-scale supercomputers~\cite{EinkemmerMoriggl2023}.

The high-dimensional Vlasov equation is a natural candidate for quantum simulation. A discretised function on \(d\) coordinates with \(N=2^n\) points per coordinate can be amplitude encoded using \(dn\) qubits rather than explicitly storing the corresponding \(N^d\) grid. Sato \emph{et al.} developed explicit Hamiltonian simulation circuits for finite-difference hyperbolic partial differential equations, whose resource requirements scale polynomially with the register width \(n\)~\cite{Sato2024}. Hu \emph{et al.} subsequently incorporated these circuit constructions into the Schr\"odingerisation framework, extending them to linear PDE discretisations with non-unitary evolution \cite{Hu2024}. Assuming the spatial resolution scales with algorithmic error (\(h=\mathcal{O}(\varepsilon)\)), their complexity analysis identifies high-dimensional regimes where a quantum advantage is possible, including \(d>5\) for the heat equation and \(d>4\) for upwind advection~\cite{Hu2024}.

Quantum algorithms have been proposed for several forms of Vlasov dynamics. Engel~\emph{et al.} formulated a Hamiltonian simulation approach for linearised Vlasov dynamics~\cite{Engel2019}. Ameri~\emph{et al.} considered the linear Vlasov equation in a Fourier--Hermite representation, including collisional effects~\cite{Ameri2023}. Toyoizumi~\emph{et al.} applied quantum singular-value transformation to Hamiltonian simulation of the linearised Vlasov--Poisson equation~\cite{Toyoizumi2024}. Miyamoto~\emph{et al.} developed a quantum algorithm for the \(6+1\)-dimensional Vlasov
description of massive-neutrino dynamics in large-scale structure
formation~\cite{Miyamoto2024}. More recently, Higuchi~\emph{et al.}
numerically evaluated Hamiltonian simulation circuits for Vlasov advection and two-stream instability within a hybrid Vlasov--Maxwell calculation~\cite{Higuchi2025}. 

These studies establish the potential of quantum algorithms for kinetic equations, but leave open the problem of constructing an explicit quantum circuit for Vlasov transport with coefficients that depend directly on the encoded phase-space coordinates. In particular, the velocity-dependent streaming term and non-linear force term require the coefficient dependence itself to be implemented coherently. This is especially relevant for hybrid quantum--classical Vlasov--Maxwell schemes~\cite{Higuchi2025}, where repeated measurement of a high-dimensional quantum state and re-encoding of updated classical fields can introduce substantial data-transfer and readout overheads.

In the present work, we consider Vlasov dynamics under a prescribed external field which arises naturally in plasma confinement and control. Einkemmer~\emph{et al.} formulated suppression of Vlasov--Poisson instabilities through optimisation of an externally applied electric field~\cite{Einkemmer2024}, while subsequent work developed analytical
external-field constructions for instability control~\cite{EinkemmerEtAl2025}. External potentials can modify particle characteristics and trapping behaviour~\cite{VelozoRuiz2024}. The prescribed-field setting provides a useful class of linear kinetic problems in which the field structure is known in advance but the resulting transport remains spatially nonuniform and nontrivial.

The contributions of this work are threefold. First, we develop and validate an explicit Schr\"odingerised circuit for variable-coefficient 1D1V Vlasov transport, including coherent evaluation of a non-linear prescribed field. Second, we extend the construction to coupled 3D3V dynamics, providing an end-to-end circuit implementation on a six-dimensional phase space. Third, we quantify the resulting logical circuit resources and examine the auxiliary \(p\)-register, including its effect on circuit cost, recovery probability, and structured state preparation. Other than state preparation and final measurement, the quantum simulation acts entirely on quantum information and requires no intermediate field evaluation or quantum--classical feedback.

\section{Theory}\label{theory}

\subsection{1D1V Vlasov equation with prescribed external field}
\label{sec:1d1v_theory}

We consider the one-dimensional-in-space, one-dimensional-in-velocity (1D1V) Vlasov transport equation with a \emph{prescribed} external acceleration field \(E(x)\):
\begin{equation}
\partial_t f(x,v,t) + v\,\partial_x f(x,v,t) + E(x)\,\partial_v f(x,v,t) = 0,
\label{eq:vlasov_1d1v_prescribed}
\end{equation}
where \(f(x,v,t)\ge 0\) denotes the phase-space density for time \(t\ge 0\). In this work, we focus on linear, externally forced dynamics without a self-consistent Poisson solver, which arise naturally in studies of confinement~\cite{Griffin-Pickering2021} and external-field control~\cite{Einkemmer2024}.

To obtain a nontrivial but structured external field, we consider the double-well potential
\begin{equation}
    V_{\rm dw}(x) = \frac{\kappa}{4}\left(x^2-a_0^2\right)^2.
\label{eq:potential_phi}
\end{equation}
The corresponding prescribed acceleration field is
\begin{equation}
    E(x) = -\frac{dV_{\rm dw}}{dx} = -\kappa\left(x^3-a_0^2x\right).
\label{eq:electric_field_E}
\end{equation}
It is convenient to rewrite this in the polynomial form
\begin{equation}
E(x)=\alpha x+\beta x^3,
\qquad
\alpha=\kappa a_0^2,
\quad
\beta=-\kappa.
\label{eq:double_well_coeffs}
\end{equation}
The prescribed field is cubic in \(x\), with coefficients \(\alpha\) and \(\beta\) that can subsequently be represented in fixed-point form in the arithmetic subroutines.

We use uniform cell-centred grids in both spatial and velocity coordinates. The quantum registers store the indices \(j\in\{0,\ldots,N_x-1\}\) and \(k\in\{0,\ldots,N_v-1\}\), corresponding to
\begin{equation}
    x_j=-\frac{L_x}{2}+\left(j+\frac12\right)\Delta x, \qquad
    v_k=-L_v+\left(k+\frac12\right)\Delta v.
\label{eq:x_v_grids}
\end{equation}
For the field-evaluation circuit, the spatial index is represented by the centred integer \(c(j)=j-N_x/2\), corresponding to \(\tilde{x}_j=\Delta x\,c(j)=x_j-\Delta x/2\). Thus, the implemented field is evaluated on a grid shifted by half a spatial cell relative to the initial distribution.

Periodic boundary conditions are imposed on both discrete transport coordinates,
\begin{equation}
    f\left(-\frac{L_x}{2},v,t\right) = f\left(\frac{L_x}{2},v,t\right), \qquad
    f(x,-L_v,t) = f(x,L_v,t),
\end{equation}
consistent with the periodic finite-difference boundary blocks. The prescribed polynomial field \(E(x)\) is evaluated at \(\tilde{x}_j\) and is not assumed to be a smooth periodic function of \(x\).

Expressing Eq.~\eqref{eq:vlasov_1d1v_prescribed} in operator form,
\begin{equation}
\partial_t f = (\mathcal{L}_x + \mathcal{L}_v)\,f,
\qquad
\mathcal{L}_x := -v\,\partial_x,
\qquad
\mathcal{L}_v := -E(x)\,\partial_v,
\label{eq:x_v_ops}
\end{equation}
using second-order Strang splitting~\cite{Strang1968,Cheng1976},
\begin{equation}
f^{n+1} \approx
e^{\frac{\Delta t}{2}\mathcal{L}_x}\;
e^{\Delta t\,\mathcal{L}_v}\;
e^{\frac{\Delta t}{2}\mathcal{L}_x}\; f^n,
\label{eq:strang_pde}
\end{equation}
where $f^n(x,v)\approx f(x,v,t_n)$ and $t_{n+1}=t_n+\Delta t$. This conveniently separates the dynamics into two transport generators that act separately on the \(x\)- and \(v\)-registers.

\subsection{Schr\"odingerisation}
\label{sec:schrodingerisation}

The Strang splitting scheme (Eq.~\eqref{eq:strang_pde}) can be implemented on \emph{explicit} quantum circuits acting on registers that encode the discretised phase-space using Hamiltonian simulation \cite{Sato2024}. In this case, however, the finite-difference scheme for the advection equation is limited to central differencing (notorious for spurious numerical oscillations), since quantum evolution must be unitary. Schr\"odingerisation \cite{Jin2024PRL} circumvents this via dimension-lifting, where one introduces an auxiliary coordinate to map a non-unitary equation to a higher-dimensional Schr\"odinger equation that is unitary, enabling dissipative equations such as the heat equation to be solved \cite{Hu2024}.

Consider a semi-discrete system first-order in time~\cite{Jin2024PRL},
\begin{equation}
    \partial_t \mathbf{f}(t)=A\,\mathbf{f}(t), 
    \label{eq:evolution_eq}
\end{equation}
where \(\mathbf{f}(t)\) is the vectorised grid representation of the phase-space density \(f\) and \(A\) is a linear evolution operator that is not necessarily skew-Hermitian. Following \cite{Jin2024PRL,Hu2024}, the semi-discrete evolution operator is first decomposed into Hermitian and anti-Hermitian parts as
\begin{equation}
    A=A_1+iA_2,
    \qquad
    A_1=\frac{A+A^\dagger}{2},
    \qquad
    A_2=\frac{A-A^\dagger}{2i},
    \label{eq:A1_A2}
\end{equation}
with both \(A_1\) and \(A_2\) Hermitian. Schr\"odingerisation introduces an auxiliary warped phase coordinate \(p\), truncated to \(p\in[-\pi R,\pi R)\) and discretised using \(N_p=2^{n_p}\) points,
\begin{equation}
    p_l = -\pi R+l\Delta p, \qquad 
    \Delta p = \frac{2\pi R}{N_p}, \qquad
    l = 0,\ldots,N_p-1.
\label{eq:p_grid}
\end{equation}
The corresponding \(n_p\) qubit register is denoted by \(P\). For \(p\ge 0\), we define
\begin{equation}
    \Phi(t,p)=e^{-p}f(t).
\label{eq:warped_solution}
\end{equation}
The initial data are extended to \(p<0\) by
\begin{equation}
    \Phi(0,p)=e^{-|p|}f(0),
\label{eq:warped_initial}
\end{equation}
or using the smoothed profile described in Appendix~\ref{app:fourier_space_loading}. Using \(\partial_p\Phi=-\Phi\), the lifted equation becomes
\begin{equation}
    \partial_t\Phi(t,p) = -A_1\partial_p\Phi(t,p) + iA_2\Phi(t,p).
\label{eq:lifted_equation}
\end{equation}

Fourier transforming in \(p\), with \(\partial_p\mapsto -i\eta\), gives
\begin{equation}
    \partial_t\widehat{\Phi}(t,\eta) = i\left(\eta A_1+A_2\right) \widehat{\Phi}(t,\eta),
\end{equation}
where
\begin{equation}
    H(\eta)=\eta A_1+A_2
\end{equation}
is Hermitian.

Let $X$ and $V$ denote the position and velocity index registers, with $n_x$ and $n_v$ qubits, respectively. We encode the discretised phase-space field at time \(t_n\) as an amplitude-encoded state,
\begin{equation}
    |\psi^n\rangle = \frac{1}{\|f^n\|_2}\sum_{j=0}^{N_x-1}\sum_{k=0}^{N_v-1}f^n_{j,k}\,
    |j\rangle_X|k\rangle_V,
\label{eq:physical_state}
\end{equation}
where
\[
    f^n_{j,k}:=f(x_j,v_k,t_n), \qquad \|f^n\|_2 = \left(\sum_{j,k}|f^n_{j,k}|^2\right)^{1/2}.
\]
The Schr\"odingerised state is obtained by augmenting this physical \(X,V\) state with the auxiliary \(P\)-register profile (Eq.~\eqref{eq:warped_solution}).
Time evolution is implemented on the lifted Schr\"odingerised state, which augments the physical phase-space grid by the auxiliary warped coordinate \(p\). We follow the explicit circuit construction of \cite{Hu2024,Sato2024}, in which finite-difference generators are decomposed into structured unitary blocks acting on encoded grid states. This way, the resulting implementation remains fully explicit and does not rely on black-box coefficient oracles.

The Schr\"odingerised implementation \cite{Hu2024} further uses an auxiliary register $P$ and, for the force step, an ancillary field register $E$ together with temporary work registers required for reversible arithmetic. The corresponding Strang-split (Eq.~\eqref{eq:strang_pde}) quantum update is
\begin{equation}
U(\Delta t) \;\approx\; U_x\!\left(\frac{\Delta t}{2}\right)\,
U_v\!\left(\Delta t\right)\,
U_x\!\left(\frac{\Delta t}{2}\right),
\label{eq:strang_quantum}
\end{equation}
where $U_x$ and $U_v$ denote the Schr\"odingerised circuit blocks for the two split substeps.

Here, we use the original auxiliary \(p\)-register \(f_p(p)=e^{-|p|}\), which admits a fixed central readout at \(p=0\). Alternative smooth profiles and structured loading strategies are discussed in Appendix~\ref{app:fourier_space_loading}.

\subsection{Velocity-dependent free-streaming circuit (Block 1)}
\label{sec:block1}

The first split substep is the free-streaming update
\begin{equation}
\partial_t f + v\,\partial_x f = 0.
\label{eq:block1_pde}
\end{equation}
We refer to the corresponding quantum implementation of \(U_x\) simply as \emph{Block 1}. Unlike \cite{Hu2024}, the advection coefficient is not a fixed scalar, but the encoded velocity itself. Consequently, the circuit angles depend explicitly on the computational basis state of the velocity register.

Let the velocity register encode the index \(k\in\{0,\dots,N_v-1\}\) of a uniform cell-centred velocity grid,
\begin{equation}
v_k = -L_v + \left(k+\tfrac{1}{2}\right)\Delta v,
\qquad
\Delta v = \frac{2L_v}{N_v}.
\label{eq:v_grid_block1}
\end{equation}
Because this grid is uniform, the encoded velocity depends affinely on the register index,
\begin{equation}
v_k = c_0 + c_1 k,
\qquad
c_0 = -L_v + \tfrac{1}{2}\Delta v,
\qquad
c_1 = \Delta v.
\label{eq:vk_affine}
\end{equation}
It follows that any phase angle proportional to \(v_k\) can be decomposed into a constant offset and a sum of bitwise contributions from the binary digits of \(k\).

If we let
\begin{equation}
k = \sum_{r=0}^{n_v-1} 2^r b_r(k), \qquad b_r(k)\in\{0,1\},
\label{eq:k_binary}
\end{equation}
where \(b_r(k)\) denotes the \(r\)-th binary digit of \(k\). The corresponding block angle takes the form
\begin{equation}
\theta(v_k) = \theta_{\mathrm{off}} + \theta_b\sum_{r=0}^{n_v-1} 2^r\,b_r(k),
\label{eq:block1_theta_decomp}
\end{equation}
where
\begin{equation}
\theta_{\mathrm{off}} = \theta_x c_0, \qquad \theta_b = \theta_x c_1, \quad
\qquad \theta_x=\frac{\Delta t}{\Delta x}.
\label{eq:block1_theta_weights}
\end{equation}
For Schr\"odingerisation, this bitwise decomposition is used with two different angle scales~\cite{Hu2024}. The \(A_1\) block, coupled to the auxiliary Fourier variable, uses
\begin{equation}
\theta_{x,1} = \frac{\theta_x}{R},
\label{eq:theta_x_1}
\end{equation}
whereas the \(A_2\) block uses
\begin{equation}
\theta_{x,2} = \theta_x .
\label{eq:theta_x_2}
\end{equation}
Here \(R\) denotes the finite truncation parameter for the auxiliary Fourier variable introduced by Schr\"odingerisation. Thus, the coefficient dependence can be implemented explicitly using one unconditional phase rotation for the offset and a collection of controlled phase rotations from the velocity qubits.

\subsubsection{Explicit circuit implementation}\label{sec:circuit_implementation}
Following~\cite{Sato2024}, the time evolution operator
\begin{equation}
\exp\!\left(i\theta\left(e^{i\lambda}s_j^- + e^{-i\lambda}s_j^+\right)\right),
\label{eq:time_evo_operator}
\end{equation}
can be implemented using
\begin{align}
W_j(\theta,\lambda)&:=B_j(\lambda)\,\mathrm{CRZ}^{\,1,\ldots,j-1}_{j}(-2\theta)\,B_j(\lambda)^\dagger, \label{eq:W_j} \\
B_j(\lambda)&:=\left(\prod_{m=1}^{j-1} \mathrm{CNOT}^{\,j}_{m}\right) P_j(-\lambda)\, H_j, \label{eq:B_j}
\end{align}
where $\lambda$ is the phase parameter used in the Bell-basis decomposition of the shift operators \(s_j^\pm\) acting as finite difference operators (here $\partial_x$ and $\partial_v$). \(\theta\) is the evolution angle due to the finite-difference coupling, where in Block~1, it is the velocity-dependent angle \(\theta(v_k)\) (Eq.~\eqref{eq:block1_theta_decomp}).

Here, an additional block \(U\) is required to enforce periodic boundary conditions
\begin{equation}
\exp\!\left(i\theta\left(e^{i\lambda}\sigma_{01}^{\otimes n_x}+e^{-i\lambda}\sigma_{10}^{\otimes n_x}\right)\right),
\label{eq:boundary_term}
\end{equation}
where \(\sigma_{01}:=|0\rangle\langle 1|\) and \(\sigma_{10}:=|1\rangle\langle 0|\). The explicit constructions of the \(W_j\) and \(U\) blocks are shown in Fig.~\ref{fig:W_j_and_U}. The reader is referred to~\cite{Hu2024,Sato2024} for details.

\begin{figure}
\centering
\[
\Qcircuit @C=1.2em @R=1.0em {
\lstick{q_0}      & \targ      & \qw        & \qw        & \qw                & \qw      & \ctrl{4} & \qw      & \qw            & \qw        & \qw        & \targ      & \qw \\
\lstick{q_1}      & \qw        & \targ      & \qw        & \qw                & \qw      & \ctrl{0} & \qw      & \qw            & \qw        & \targ      & \qw        & \qw \\
\lstick{\vdots}   &            &            &            &                    &          &          &          &                &            &            &            &     \\
\lstick{q_{j-1}}  & \qw        & \qw        & \targ      & \qw                & \qw      & \ctrl{0} & \qw      & \qw            & \targ      & \qw        & \qw        & \qw \\
\lstick{q_j}      & \ctrl{-4}  & \ctrl{-3}  & \ctrl{-1}  & \gate{P(\lambda)}  & \gate{H} & \gate{R_Z(\theta)} & \gate{H} & \gate{P(-\lambda)} & \ctrl{-1} & \ctrl{-3} & \ctrl{-4} & \qw \\
\lstick{\vdots}   &            &            &            &                    &          &          &          &                &            &            &            &     \\
\lstick{q_{n-1}}  & \qw        & \qw        & \qw        & \qw                & \qw      & \qw      & \qw      & \qw            & \qw        & \qw        & \qw        & \qw
}
\]
(a)
\vspace{0.5cm}
\[
\Qcircuit @C=1.2em @R=1.0em {
\lstick{q_0}      & \targ      & \qw        & \qw        & \qw                &  \qw   & \ctrlo{1} &  \qw  & \qw            & \qw        & \qw        & \targ      & \qw \\
\lstick{q_1}      & \qw        & \targ      & \qw        & \qw                &  \qw   & \ctrlo{2} &  \qw  & \qw            & \qw        & \targ      & \qw        & \qw \\
\lstick{\vdots}   &            &            &            &                    &          &          &          &                &            &            &            &     \\
\lstick{q_{j-1}}  & \qw        & \qw        & \targ      & \qw                &  \qw   & \ctrlo{1} & \qw  & \qw            & \targ      & \qw        & \qw        & \qw \\
\lstick{q_j}      & \ctrl{-4}  & \ctrl{-3}  & \ctrl{-1}  & \gate{P(\lambda)}  & \gate{H} & \gate{R_Z(\theta)} & \gate{H} & \gate{P(-\lambda)} & \ctrl{-1} & \ctrl{-3} & \ctrl{-4} & \qw \\
\lstick{\vdots}   &            &            &            &                    &          &          &          &                &            &            &            &     \\
\lstick{q_{n-1}}  & \qw        & \qw        & \qw        & \qw                & \qw      & \qw      & \qw      & \qw            & \qw        & \qw        & \qw        & \qw
}
\]
(b)
\caption{Generic coefficient-dependent (a) $W_j(\theta,\lambda)$ block and (b) $U(\theta,\lambda)$ boundary block. Adapted from~\cite{Hu2024,Sato2024}.}
\label{fig:W_j_and_U}
\end{figure}

\begin{figure}
\centering
\[
\Qcircuit @C=1.0em @R=1.1em {
\lstick{x_j} 
    & \gate{\theta_{\rm off}} 
    & \gate{2^0\theta_b} 
    & \gate{2^1\theta_b} 
    & \qw    & \qw
    & \gate{2^{n_v-1}\theta_b}
    & \qw
    & \push{\rule{1em}{0em}}
    & \qw
    & \gate{-2\theta_{\rm off}}
    & \gate{-2^1\theta_b}
    & \gate{-2^2\theta_b}
    & \qw    
    & \qw
    & \gate{-2^{n_v-1}\theta_b}
    & \qw \\
\lstick{v_0} 
    & \qw 
    & \ctrl{-1} 
    & \qw 
    & \qw
    & \qw
    & \qw    & \qw
    & \push{\rule{1em}{0em}}
    & \qw    & \qw
    & \ctrl{0}
    & \qw
    & \qw
    & \qw
    & \qw
    & \qw \\
\lstick{v_1} 
    & \qw 
    & \qw 
    & \ctrl{-2} 
    & \qw
    & \qw
    & \qw    & \qw
    & \push{\rule{1em}{0em}}
    & \qw    
    & \qw
    & \qw
    & \ctrl{0}
    & \qw
    & \qw
    & \qw
    & \qw
    & \\
\lstick{\vdots}
    & & & & & \lstick{\ddots}
    & & & & & & & & & \lstick{\ddots}
    \\
\lstick{v_{n_v-2}} 
    & & & & & & & & & & & & & &
    & \ctrl{0}
    & \qw
    \\
\lstick{v_{n_v-1}} 
    & \qw 
    & \qw 
    & \qw 
    & \qw    & \qw
    & \ctrl{-5}
    & \qw
    & \push{\rule{1em}{0em}} 
    & \qw
    & \ctrlo{-5}
    & \ctrlo{-5}
    & \ctrlo{-5}
    & \qw    & \qw
    & \ctrlo{-5}
    & \qw
}
\]
\caption{Velocity-dependent free-streaming block \(vW_j\). The phase labels indicate the elementary controlled-\(R_Z\) angles arising from the decomposition of \(W_j\). The left portion implements the signed velocity contribution for the centred velocity encoding, while the portion after the gap applies the additional \(|v|\)-dependent correction required by the \(A_1\) term in Eq.~\eqref{eq:A1_A2}.}

\label{fig:vWj-block}
\end{figure}
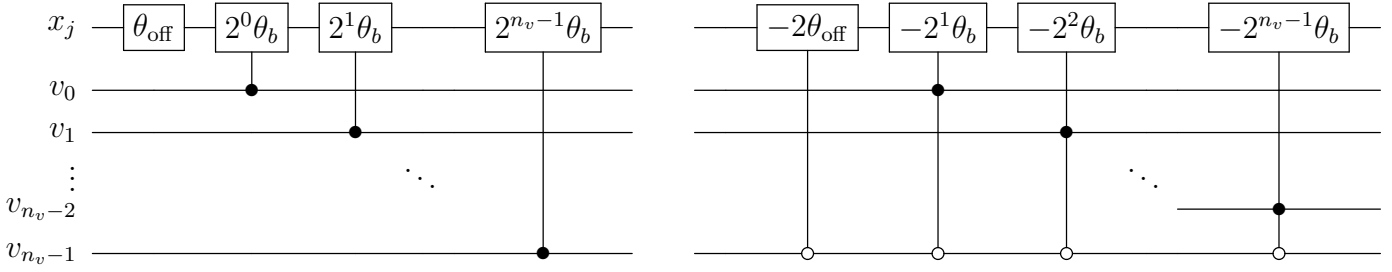

To incorporate the encoded velocity dependence, the evolution angle \(\theta\) in each \(W_j\) block is replaced by the bitwise decomposition in Eq.~\eqref{eq:block1_theta_decomp}. The resulting velocity-controlled free-streaming block, denoted \(vW_j\), is shown in Fig.~\ref{fig:vWj-block}. Its phase labels \(\theta_0,2^0\theta_b,2^1\theta_b,\ldots\) represent the elementary controlled-\(R_Z\) rotations generated by the binary expansion of the velocity index. The left portion of the circuit implements the signed contribution proportional to \(v_k\), while the separated portion applies the additional \(|v_k|\)-dependent correction required by the \(A_1\) part of the Schr\"odingerised generator in Eq.~\eqref{eq:A1_A2}. The boundary unitary \(vU\) is modified in the same manner.

The free-streaming update is assembled from two composite operators, \(vV_1\) and \(vV_2\), corresponding respectively to the \(A_1\) and \(A_2\) components of the Schr\"odingerised generator. The block \(vV_1\) is constructed from the bulk block \(vW_j(\lambda=0)\) together with the boundary block \(vU(\lambda=0)\). The companion block \(vV_2\) has the same circuit structure, but uses the phase choices \(\lambda=-\pi/2\) for the bulk terms and \(\lambda=\pi/2\) for the boundary term. In both cases, the phase rotations \(\theta_0\) and \(\theta_b\) act on the auxiliary phase register \(P\) and are controlled by the velocity register \(V\), thereby coupling the lifted auxiliary dynamics to the spatial transport.

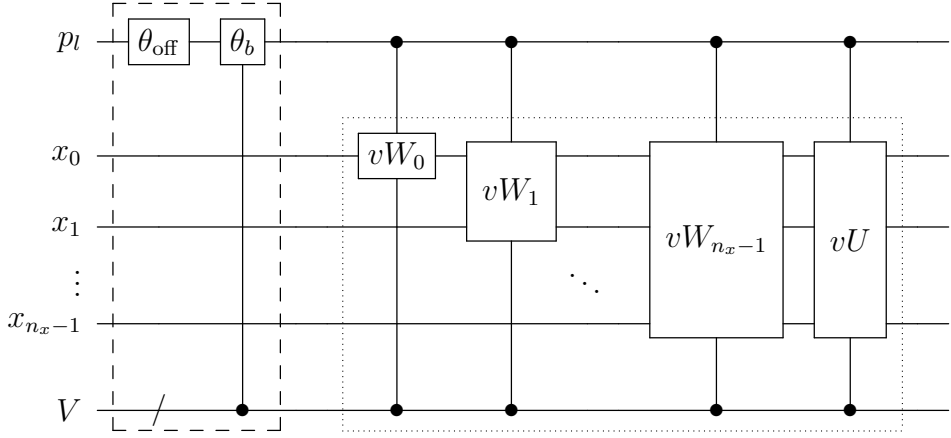
\begin{figure}
\centering
\[
\Qcircuit @C=1.0em @R=1.1em {
\lstick{p_l} & \gate{\theta_{\rm off}} & \gate{\theta_b}
             & \qw
             & \qw
             & \ctrl{2}
             & \ctrl{2}
             & \qw & \qw
             & \ctrl{2}
             & \ctrl{2}
             & \qw 
             & \qw 
             \\
             \\
\lstick{x_0} & \qw 
             & \qw
             & \qw
             & \qw
             & \gate{vW_0}
             & \multigate{1}{vW_1}
             & \qw & \qw
             & \multigate{3}{vW_{n_x-1}}
             & \multigate{3}{vU}
             & \qw 
             & \qw \\
\lstick{x_1} & \qw & \qw
             & \qw
             & \qw
             & \qw
             & \ghost{vW_1}
             & \qw & \qw
             & \ghost{vW_{n_x-1}}
             & \ghost{vU}
             & \qw
             & \qw \\
\lstick{\vdots}  
             & & & & & & & & 
             \lstick{\ddots}
            \\
\lstick{x_{n_x-1}}  
             & \qw & \qw
             & \qw
             & \qw
             & \qw
             & \qw
             & \qw & \qw
             & \ghost{vW_{n_x-1}}
             & \ghost{vU}
             & \qw
             & \qw \\
             \\
\lstick{V}   & \qw {/}
             & \ctrl{-7}
             & \qw
             & \qw 
             & \ctrl{-5}
             & \ctrl{-4}
             & \qw & \qw
             & \ctrl{-2}
             & \ctrl{-2}
             & \qw
             & \qw
\gategroup{1}{2}{8}{3}{1em}{--}
\gategroup{3}{6}{8}{11}{1em}{..}
}
\]
\caption{\(vV_1\) block for \(A_1\) term. The dashed region applies the phase rotations \(\theta_0\) and \(\theta_b\) on the \(p\)-register, controlled by the velocity register, while the remaining controlled \(vW_j\) and \(vU\) blocks implement the bulk and boundary finite-difference couplings. The dotted region indicates the corresponding \(vV_2\) block for the \(A_2\) term, which has the same layout but uses phase selections \(vW_j(\lambda=-\pi/2)\) and \(vU(\lambda=\pi/2)\).}
\label{fig:vV1_vV2}
\end{figure}

Figure~\ref{fig:vV1_vV2} displays the circuit structure of \(vV_1\). The dashed region applies the velocity-controlled phase rotations on the auxiliary \(P\)-register, while the remaining controlled \(vW_j\) and \(vU\) blocks realise the bulk and boundary finite-difference couplings in the spatial register. The dotted region highlights the corresponding \(vV_2\) block. Its architecture is identical, differing only in the phase selections \(vW_j(\lambda=-\pi/2)\) and \(vU(\lambda=\pi/2)\) required for the \(A_2\) contribution.
 
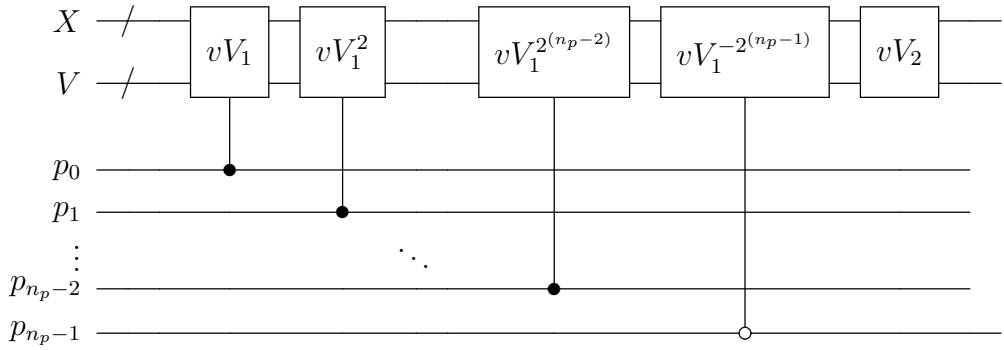
\begin{figure}
\centering
\[
\Qcircuit @C=1.0em @R=1.1em {
\lstick{X}  & \qw {/}   & \qw   
            & \multigate{1}{vV_1}
            & \multigate{1}{vV_1^{2}}
            & \qw   & \qw   
            & \multigate{1}{vV_1^{2^{(n_p-2)}}}
            & \multigate{1}{vV_1^{-2^{(n_p-1)}}}
            & \multigate{1}{vV_2}
            & \qw   & \qw
            \\
\lstick{V}  & \qw {/}   & \qw   
            & \ghost{vV_1}
            & \ghost{vV_1^{2}}
            & \qw   & \qw   
            & \ghost{vV_1^{2^{(n_p-2)}}}
            & \ghost{vV_1^{-2^{(n_p-1)}}}
            & \ghost{vV_2}
            & \qw   & \qw
            \\ \\
\lstick{p_0} & \qw      & \qw   
             & \ctrl{-2}
             & \qw
             & \qw  & \qw
             & \qw  & \qw   & \qw
             & \qw \\
\lstick{p_1} & \qw      & \qw  
             & \qw
             & \ctrl{-3}
             & \qw  & \qw
             & \qw  & \qw   & \qw
             & \qw \\
\lstick{\vdots}  
             & & & & & &
             \lstick{\ddots}
             \\
\lstick{p_{n_p-2}}      & \qw   & \qw   & \qw
             & \qw      & \qw   & \qw
             & \ctrl{-5}
             & \qw
             & \qw 
             & \qw \\
\lstick{p_{n_p-1}}      & \qw   & \qw   & \qw
             & \qw      & \qw   & \qw   & \qw
             & \ctrlo{-6}
             & \qw
             & \qw 
             & \qw \\
}
\]
\caption{Construction of Block 1 from binary-controlled powers of \(vV_1\), followed by a single \(vV_2\) block. Each \(P\)-register qubit controls the corresponding power \(vV_1^{2^l}\), and the most significant qubit controls the adjoint unitaries.}
\label{fig:block_1}
\end{figure}

Block 1 is obtained by combining binary-controlled powers of \(vV_1\) with a single application of \(vV_2\), as shown in Fig.~\ref{fig:block_1}. For each \(l=0,1,\dots,n_p-1\), the \(l\)-th qubit of the auxiliary phase register \(P\) controls the power \(vV_1^{2^l}\), which encodes the dependence on the discretised auxiliary variable through its binary expansion. The most significant qubit controls the adjoint branch. As noted in~\cite{Hu2024}, the total number of elementary \(vV_1\) applications scales as \(\sum_{l=0}^{n_p-1}2^l\), and therefore grows exponentially with the size of the auxiliary phase register.

Although \(n_p\) can be chosen independently of the physical phase-space resolution, its appropriate value depends on the problem, the auxiliary-domain parameter \(R\), and the desired accuracy. Increasing \(n_p\) could substantially increase circuit resources without necessarily improving solution accuracy (Secs.~\ref{sec:results_3d3v} and \ref{sec:3d3v_resources}). The associated trade-offs in auxiliary state preparation and
readout are further examined in Appendix~\ref{app:fourier_space_loading}.

\subsection{Field-dependent acceleration circuit (Block 2)}
\label{sec:block2}

The second split substep is the field-driven acceleration evolution
\begin{equation}
\partial_t f + E(x)\,\partial_v f = 0.
\label{eq:block2_pde}
\end{equation}
We refer to the quantum implementation of \(U_v\) as \emph{Block 2}. As in Block 1, the Schr\"odingerised construction is based on the discretised transport generator for the split substep, but the advection coefficient is now the position-dependent field \(E(x)\). 

For the cubic double-well field in Eq.~\eqref{eq:double_well_coeffs}, the required phase angles are not affine in the binary register index, so their implementation requires explicit arithmetic to evaluate the field dependence before applying the controlled rotations. Using the spatial grid defined in Eq.~\eqref{eq:x_v_grids}, the field value corresponding to the computational basis state \(\ket{j}_X\) must first be constructed coherently on an auxiliary arithmetic register through the reversible map
\begin{equation}
    \ket{j}_X\ket{0}_E \;\mapsto\; \ket{j}_X\ket{E_{\mathrm{fp}}(j)}_E, \qquad 
    \frac{E_{\mathrm{fp}}(j)}{S} \approx E(\tilde{x}_j).
\label{eq:E_register_map}
\end{equation}

Schematically, the principal data registers for coherent field evaluation are
\begin{equation}
\ket{j}_X
\ket{0}_{c}
\ket{0}_{c_{\mathrm{abs}}}
\ket{0}_{\mathrm{xcopy}}
\ket{0}_{\mathrm{tmp}}
\ket{0}_{\mathrm{sq}}
\ket{0}_{\mathrm{cube}}
\ket{0}_{E}.
\label{eq:block2_register_layout}
\end{equation}
Here \(X\) is the position register, \(c\) is the centred signed register derived from the encoded position value, \(c_{\mathrm{abs}}\) stores its unsigned magnitude, \(\mathrm{xcopy}\), \(\mathrm{tmp}\), \(\mathrm{sq}\), and \(\mathrm{cube}\) are work registers used in the polynomial evaluation, and \(E\) stores the fixed-point field value \(E_{\mathrm{fp}}(j)\) corresponding to \(E(\tilde{x}_j)\). Additional carry and sign ancillas required by the reversible arithmetic routines are omitted from Eq.~\eqref{eq:block2_register_layout} for clarity.

A conservative choice of the output-register width \(n_E\) is obtained by bounding the maximum magnitude of the fixed-point field accumulator. Let \(\alpha_{\mathrm{fp}}\) and \(\beta_{\mathrm{fp}}\) denote the signed fixed-point integer representations of the real coefficients \(\alpha\) and \(\beta\). If the centred position register ranges over integers of magnitude at most \(2^{n_x-1}\), then the fixed-point field accumulator satisfies
\begin{equation}
|E_{\mathrm{fp}}| \le M_E,
\qquad
M_E :=
|\alpha_{\mathrm{fp}}|\,2^{n_x-1}
+
|\beta_{\mathrm{fp}}|\,2^{3(n_x-1)}.
\label{eq:ME_bound}
\end{equation}
A sufficient two's-complement width for the output register is therefore
\begin{equation}
n_E=
\max\!\left(
3n_x+1,\;
\left\lceil \log_2(M_E+1)\right\rceil + 1
\right),
\label{eq:nE_width}
\end{equation}
where the second term provides enough bits to represent any signed integer of magnitude at most \(M_E\), and the first term enforces a structural lower bound compatible with the signed cubic workspace.

For the present cubic double-well field, the additional arithmetic required by Block 2 does not introduce an exponential qubit overhead in the grid size.
\begin{lemma}
\label{lem:qubit_scaling}
For the cubic double-well field \(E(x)=\alpha x+\beta x^3\) with fixed-point coefficients of bounded precision, the auxiliary arithmetic workspace of Block 2 scales linearly with \(n_x\), and hence as \(O(\log N_x)\) in the spatial grid size \(N_x=2^{n_x}\).
\end{lemma}

\begin{proof}
The coherent field-evaluation subroutine uses principal registers of widths \(n_x\), \(n_x\), \(n_x\), \(n_x\), \(2n_x\), \(4n_x\), and \(n_E\), corresponding respectively to \(c\), \(c_{\mathrm{abs}}\), \(\mathrm{xcopy}\), \(\mathrm{tmp}\), \(\mathrm{sq}\), \(\mathrm{cube}\), and \(E\). In addition, the reversible signed-arithmetic routines require carry and sign ancillas of total width \(O(n_x+n_E)\). Hence the total auxiliary workspace is \(O(n_x+n_E)\). From Eq.~\eqref{eq:nE_width}, for bounded fixed-point precision one has \(n_E=3n_x+O(1)\). Therefore the total auxiliary workspace is \(O(n_x)\), which is \(O(\log N_x)\) since \(N_x=2^{n_x}\).
\end{proof}

The linear workspace scaling is not specific to the cubic field of choice here. 
\begin{remark}
\label{rem:qubit_scaling}
More generally, if \(E(x)\) is a polynomial of fixed degree \(d\) with coefficients represented at bounded fixed-point precision, then coherent evaluation of \(E(\tilde{x}_j)\) still requires only \(O(n_x)\) auxiliary qubits. The number of monomial terms is constant, and the largest monomial \(c(j)^d\) requires only \(O(d\,n_x)=O(n_x)\) bits when \(d\) is independent of \(n_x\). Hence the total arithmetic workspace remains linear in \(n_x\), or equivalently logarithmic in \(N_x=2^{n_x}\).
\end{remark}

\subsubsection{Coherent construction of the auxiliary field register}
\label{sec:field_register}

We coherently prepare a field register whose contents encode the fixed-point field value used in the acceleration step. First, the unsigned grid index \(j\) is converted into the centred signed integer
\begin{equation}
    c(j)=j-2^{n_x-1},
\label{eq:centered_integer}
\end{equation}
stored in two's-complement form. 

The centred integer corresponds to the field-evaluation coordinate
\begin{equation}
    \tilde{x}_j=\Delta x\,c(j) =x_j-\frac{\Delta x}{2}.
\label{eq:centered_x}
\end{equation}
Hence, the cubic double-well field evaluated by the circuit is
\begin{equation}
    E(\tilde{x}_j) =\alpha\Delta x\,c(j)+\beta(\Delta x)^3c(j)^3.
\label{eq:centered_field}
\end{equation}


We use a signed fixed-point representation with \(s\) fractional bits and scale factor \(S:=2^s\). The real coefficients of \(c(j)\) and \(c(j)^3\) are encoded as fixed-point integers by
\begin{equation}
\alpha_{\mathrm{fp}} = \operatorname{round}(S\alpha\Delta x),
\qquad \beta_{\mathrm{fp}} = \operatorname{round}(S\beta(\Delta x)^3).
\end{equation}
The target integer stored in the output field register is therefore
\begin{equation}
E_{\mathrm{fp}}(j) = \alpha_{\mathrm{fp}}\,c(j) + \beta_{\mathrm{fp}}\,c(j)^3 .
\label{eq:block2_Efp}
\end{equation}

\begin{figure}[htbp]
\centering
\[
\Qcircuit @C=0.7em @R=1.0em {
\lstick{\makebox[1.2cm][l]{$\ket{j}_X$}}
    & \ctrl{3}
    & \qw
    & \qw
    & \qw
    & \qw   &\qw    &\qw   &\qw    &\qw    & \qw   &\qw    &\qw   &\qw
    & \qw   &\qw    &\qw    &\qw   
    &\rstick{\ket{j}_X}
    \\  \\ \\
\lstick{\makebox[1.2cm][l]{$\ket{0}_c$}}
    & \targ
    & \gate{\mathrm{ctr}}
    & \ctrl{1}
    & \sgate{\mathrm{sign}}{1}
    & \qw
    & \qw   &\qw    & \qw   &\qw    &\qw    & \qw   &\qw
    & \sgate{\mathrm{sign}}{5}
    & \sgate{\mathrm{acc}}{6}
    & \qw   &\qw    &\qw
    &\rstick{\ket{c}_c}
    \\
\lstick{\makebox[1.2cm][l]{$\ket{0}_{c_{\mathrm{abs}}}$}}
    & \qw    &\qw
    & \targ
    & \gate{\mathrm{tc}}
    & \ctrl{1}
    & \qw
    & \qw
    & \qw
    & \qw
    & \qw
    & \qw
    & \qw
    & \qw
    & \qw    &\qw    &\qw    & \qw 
    & \rstick{\ket{|c|}_{c_{\mathrm{abs}}}}
    \\
\lstick{\makebox[1.2cm][l]{$\ket{0}_{\mathrm{xcopy}}$}}
    & \qw    &\qw
    & \qw
    & \qw
    & \targ
    & \ctrl{1}
    & \multigate{2}{\mathrm{mul}}
    & \ctrl{1}
    & \multigate{3}{\mathrm{mul}}
    & \ctrl{1}
    & \multigate{2}{\mathrm{mul}^\dag}
    & \ctrl{1}
    & \qw   & \qw   &\qw   & \qw   &\qw
    & \rstick{\ket{0}_{\mathrm{xcopy}}}
    \\
\lstick{\makebox[1.2cm][l]{$\ket{0}_{\mathrm{tmp}}$}}
    & \qw    &\qw
    & \qw
    & \qw
    & \qw
    & \targ
    & \ghost{\mathrm{mul}}
    & \targ
    & \ghost{\mathrm{mul}}
    & \targ
    & \ghost{\mathrm{mul}^\dag}
    & \targ
    & \qw   & \qw   &\qw    & \qw   &\qw
    & \rstick{\ket{0}_{\mathrm{tmp}}}
    \\
\lstick{\makebox[1.2cm][l]{$\ket{0}_{\mathrm{sq}}$}}
    & \qw    &\qw
    & \qw
    & \qw
    & \qw
    & \qw
    & \ghost{\mathrm{mul}}
    & \qw 
    & \ghost{\mathrm{mul}}
    & \qw
    & \ghost{\mathrm{mul}^\dag}
    & \qw   &\qw   &\qw    &\qw    &\qw    &\qw
    & \rstick{\ket{0}_{\mathrm{sq}}}
    \\
\lstick{\makebox[1.2cm][l]{$\ket{0}_{\mathrm{cube}}$}}
    & \qw    &\qw   & \qw   &\qw   & \qw   &\qw   & \qw   &\qw
    & \ghost{\mathrm{mul}}
    & \qw
    & \qw
    & \qw
    & \gate{\mathrm{tc}}
    & \qw   
    & \sgate{\mathrm{acc}}{1}
    &\qw
    & \qw 
    & \rstick{\ket{c^3}_{\mathrm{cube}}}
    \\
\lstick{\makebox[1.2cm][l]{$\ket{0}_{E}$}}
    & \qw   & \qw   &\qw   & \qw   &\qw   & \qw   &\qw   & \qw   &\qw   & \qw   &\qw
    & \qw    &\qw
    & \gate{\mathrm{acc}}
    & \gate{\mathrm{acc}}
    & \qw
    & \qw 
    & \rstick{\ket{E}_{E}}
\gategroup{6}{9}{7}{9}{1em}{..}
\gategroup{6}{13}{7}{13}{1em}{..}
}
\]
\caption{High-level workflow of Block 2. Here, \(\mathrm{ctr}\) denotes the MSB flip that yields the centred signed register \(c\), \(\mathrm{tc}\) denotes controlled two's complement negation, \(\mathrm{mul}\) denotes reversible multiplication, \(\mathrm{sign}\) denotes sign restoration, and \(\mathrm{acc}\) denotes weighted accumulation into \(E\). All \(\mathrm{CNOT}\) gates shown denote bitwise \(\mathrm{CNOT}\) ladders and the dotted box indicates ladders applied in reverse order for uncomputation. Additional work and ancilla registers are omitted for clarity.}
\label{fig:block2_workflow}
\end{figure}
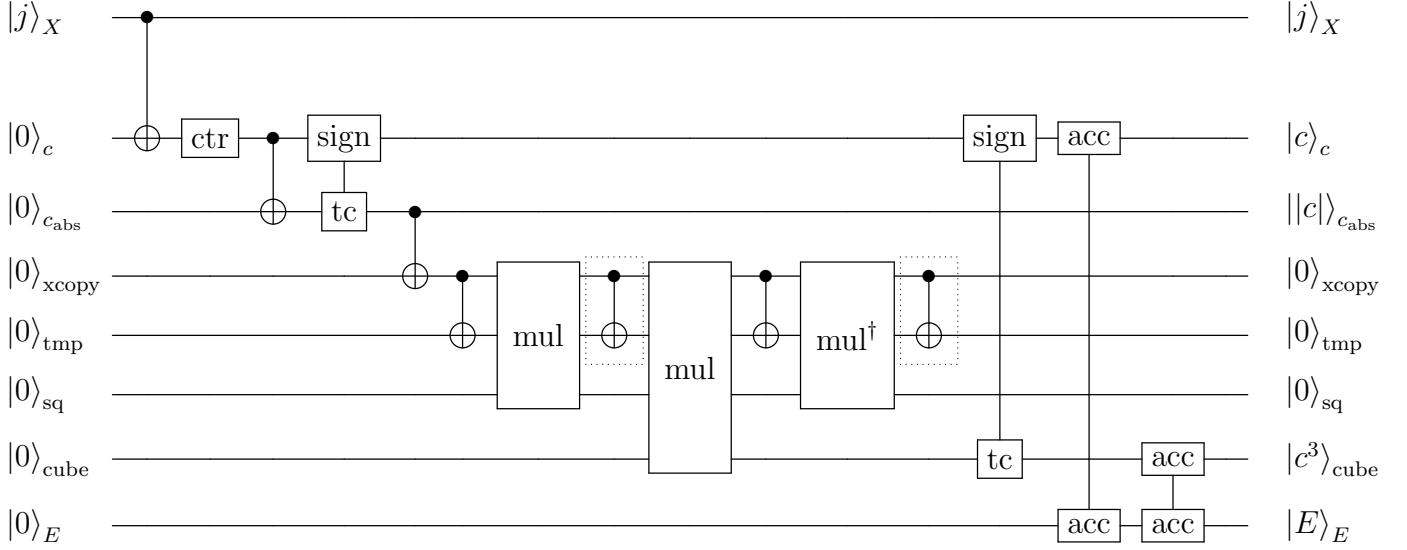

Figure~\ref{fig:block2_workflow} summarises the auxiliary field-register
construction as high-level reversible data flow into the output field register
\(E_{\mathrm{fp}}\). Additional gate-level arithmetic details can be found in
Appendix~\ref{app:block2_arithmetic}.

Here, we outline step-by-step the reversible construction of the \(E_{\mathrm{fp}}\) register:

\paragraph{Step 1: Preparation of the centred signed register.}
Starting from the encoded position register, the spatial index is reversibly mapped to a register storing the centred signed integer \(c(j)\):
\begin{equation}
\ket{j}_X\ket{0}_{c}
\;\mapsto\;
\ket{j}_X\ket{c(j)}_{c}.
\label{eq:block2_step1}
\end{equation}

\paragraph{Step 2: Magnitude extraction.}
The unsigned magnitude register is then prepared from the signed register \(c\):
\begin{equation}
\ket{c}_{c}\ket{0}_{c_{\mathrm{abs}}}
\;\mapsto\;
\ket{c}_{c}\ket{|c|}_{c_{\mathrm{abs}}}.
\label{eq:block2_step2}
\end{equation}

\paragraph{Step 3: Square and cubic workspaces.}
Using reversible multiplication, the magnitude register is promoted to quadratic and cubic work registers. First, the square register is constructed as
\begin{equation}
\ket{|c|}_{c_{\mathrm{abs}}}\ket{0}_{\mathrm{sq}}
\;\mapsto\;
\ket{|c|}_{c_{\mathrm{abs}}}\ket{|c|^2}_{\mathrm{sq}}.
\label{eq:block2_step3a}
\end{equation}
This intermediate result is then used to form the cubic register,
\begin{equation}
\ket{|c|}_{c_{\mathrm{abs}}}\ket{|c|^2}_{\mathrm{sq}}\ket{0}_{\mathrm{cube}}
\;\mapsto\;
\ket{|c|}_{c_{\mathrm{abs}}}\ket{|c|^2}_{\mathrm{sq}}\ket{|c|^3}_{\mathrm{cube}}.
\label{eq:block2_step3b}
\end{equation}
Once the cubic term has been obtained, the square register \(\mathrm{sq}\) is uncomputed to \(\ket{0}\).

\paragraph{Step 4: Sign restoration for the cubic term.}
The cubic register is then conditionally negated according to the sign of \(c\), yielding a signed representation of \(c^3\):
\begin{equation}
\ket{c}_{c}\ket{|c|^3}_{\mathrm{cube}}
\;\mapsto\;
\ket{c}_{c}\ket{c^3}_{\mathrm{cube}}.
\label{eq:block2_step4}
\end{equation}

\paragraph{Step 5: Weighted accumulation into the field register.}
Let \(\alpha_{\mathrm{fp}}\) and \(\beta_{\mathrm{fp}}\) denote the fixed-point integer representations of the coefficients \(\alpha\) and \(\beta\). The output register \(E\) is then assembled by reversibly accumulating the weighted linear and cubic contributions,
\begin{equation}
\ket{c}_{c}\ket{c^3}_{\mathrm{cube}}\ket{0}_{E}
\;\mapsto\;
\ket{c}_{c}\ket{c^3}_{\mathrm{cube}}
\ket{\alpha_{\mathrm{fp}}c+\beta_{\mathrm{fp}}c^3}_{E}.
\label{eq:block2_step5}
\end{equation}
Thus the field register stores the fixed-point \(E_{\mathrm{fp}}\) which represents the discretised field value associated with the basis state \(\ket{j}_X\).

To form the weighted field value \(E_{\mathrm{fp}}=\alpha_{\mathrm{fp}}c+\beta_{\mathrm{fp}}c^3\), the circuit must apply the fixed coefficients \(\alpha_{\mathrm{fp}}\) and \(\beta_{\mathrm{fp}}\) to the signed source registers \(c\) and \(c^3\). Here, this is done through \emph{controlled constant-addition blocks}, namely reversible operations that, conditioned on a single source bit, add the corresponding classical shifted constant into the output register \(E\) while leaving the source register unchanged. 

\subsubsection{Resource estimates for polynomial field accumulation}
\label{app:block2_arithmetic_resources}

The number of controlled constant-addition blocks required depends on the widths of the signed source registers.

\begin{lemma}[Controlled-addition count for polynomial field accumulation]
\label{lem:accumulation_cost_scales_nx}
Suppose the field is written as
\begin{equation}
E(x)=\sum_{m\in S} a_m x^m,
\end{equation}
where \(S\subset \mathbb N\) is a fixed finite set of nonzero polynomial exponents. Assume that, for each \(m\in S\), the signed value \(x^m\) is stored in a register of width \(m n_x\), and that the contribution \(a_m x^m\) is accumulated into the output register using the two's-complement bits of this source register. Then the number of controlled constant-addition blocks used by this accumulation step is
\begin{equation}
n_x\sum_{m\in S} m.
\end{equation}
In particular, if \(S\) is independent of \(n_x\), then the number of controlled constant-addition blocks scales as \(O(n_x)\).
\end{lemma}

\begin{proof}
For a given exponent \(m\in S\), the quantity \(x^m\) is stored in an \(mn_x\)-bit signed source register. In a bitwise two's complement accumulation, one controlled constant-addition block is used for each bit of the source register. The \(m n_x-1\) non-sign bits each contribute a positively weighted shifted constant, while the sign bit contributes the corresponding negatively weighted shifted constant. Therefore the contribution \(a_m x^m\) requires
\begin{equation}
(m n_x-1)+1 = m n_x
\end{equation}
controlled constant-addition blocks. Summing over all nonzero polynomial terms gives
\begin{equation}
\sum_{m\in S} m n_x = n_x\sum_{m\in S} m.
\end{equation}
If \(S\) is fixed independently of \(n_x\), then \(\sum_{m\in S}m\) is a
constant, so the accumulation cost is \(O(n_x)\).
\end{proof}

For the cubic double-well field \(E(x)=\alpha x+\beta x^3\), we have \(S=\{1,3\}\). Hence the weighted accumulation step uses \(n_x(1+3)=4n_x\) controlled constant-addition blocks.

\bigskip

We next estimate the depth cost of the auxiliary field-register construction
used in Block~2.

\begin{lemma}[Depth estimate for the auxiliary field-register construction]
\label{lem:field_register_depth}
For the cubic auxiliary field \(E(x)=\alpha x+\beta x^3\), let \(D_{\mathrm{mul}}(m)\) denote the depth of the reversible multiplier used on \(m\)-bit inputs. Assume that each controlled constant-addition block into the \(n_E\)-qubit output register has depth \(O(n_E)\). Then the depth \(D_E\) of the coherent field-register construction satisfies
\begin{equation}
D_E \le 2D_{\mathrm{mul}}(n_x) + D_{\mathrm{mul}}(2n_x) + O(n_x n_E).
\end{equation}
In particular, since \(n_E=3n_x+O(1)\), this gives
\begin{equation}
D_E \le 2D_{\mathrm{mul}}(n_x) + D_{\mathrm{mul}}(2n_x) + O(n_x^2).
\end{equation}
Hence the auxiliary field-register construction has polynomial depth whenever the chosen reversible multiplier family has polynomial depth.
\end{lemma}

\begin{proof}
The preprocessing and sign-handling steps have depth \(O(n_x)\), which is absorbed into the \(O(n_x n_E)\) accumulation term. The cubic contribution is formed by computing \(|c|^2\), then computing \(|c|^3\), and finally uncomputing the intermediate square register. This gives the multiplier-depth contribution
\begin{equation}
2D_{\mathrm{mul}}(n_x) + D_{\mathrm{mul}}(2n_x).
\end{equation}
Here the second multiplication is bounded by \(D_{\mathrm{mul}}(2n_x)\) because it multiplies the \(n_x\)-bit magnitude \(|c|\) by the \(2n_x\)-bit square \(|c|^2\).

The accumulation stage adds \(\alpha_{\mathrm{fp}}c+\beta_{\mathrm{fp}}c^3\) into \(E\) by scanning the two's-complement bits of the signed source registers. Since the linear source register has width \(n_x\) and the cubic source register has width \(3n_x\), Lemma~\ref{lem:accumulation_cost_scales_nx} gives \(4n_x\) controlled constant-addition blocks. Each such block has depth \(O(n_E)\), so the total accumulation depth is \(O(n_x n_E)\).
\end{proof}

\subsubsection{Field-controlled acceleration block}
\label{sec:block2_field_controlled_acceleration}

We now assemble the Schr\"odingerised circuit for the field-driven acceleration step Eq.~\eqref{eq:block2_pde} as an analogue of the velocity-controlled free-streaming block in Sec.~\ref{sec:block1}, except that the finite difference now acts on the velocity register \(V\), and the advection coefficient is the coherently computed field value \(E(\tilde{x}_j)\) rather than the encoded velocity \(v_k\).

For the field-controlled rotations, the coefficient has already been computed into the \(n_E\)-qubit field register. The field-dependent angle is
\begin{equation}
\theta_E(j) \approx \theta_sE_{\rm fp}(j), \quad \theta_s = \frac{\theta_v}{S}, 
\quad \theta_v = \frac{\Delta t}{\Delta v}.
\end{equation}
Substituting the two's-complement expansion gives
\begin{equation}
\theta_E(j) = \theta_s
\left(\sum_{r=0}^{n_E-2} 2^r e_r(j)-2^{n_E-1} e_{n_E-1}(j)\right), \quad e_r \in \{0,1\},
\end{equation}
where \(e_{n_E-1}(j)\) is the sign bit. Note that unlike Eq.~\eqref{eq:block1_theta_weights} in Block~1, the field-register decomposition has no affine offset term, i.e. \(\theta_{\rm off}^{(E)}=0\). Similarly, the \(A_1\) block, coupled to the auxiliary Fourier variable, uses
\begin{equation}
\theta_{v,1} = \frac{\theta_v}{R},
\label{eq:theta_v_1}
\end{equation}
whereas the \(A_2\) block uses
\begin{equation}
\theta_{v,2} = \theta_v .
\label{eq:theta_v_2}
\end{equation}

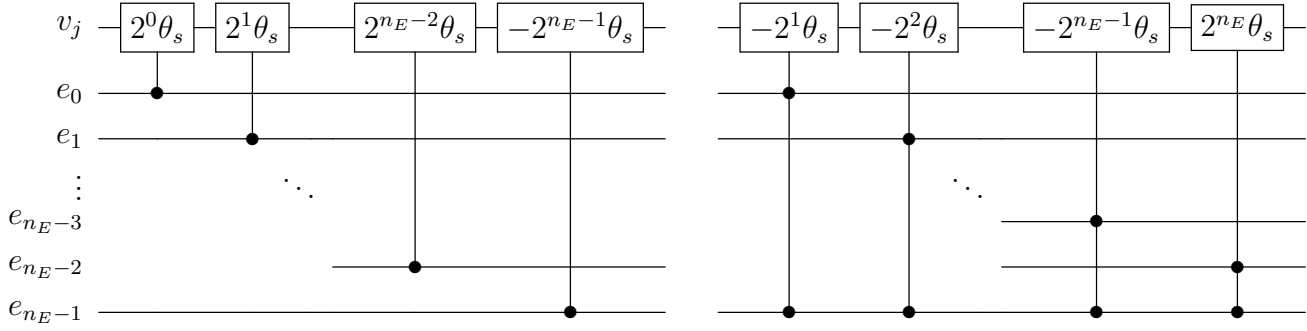
\begin{figure}
\centering
\[
\Qcircuit @C=0.7em @R=1.2em {
\lstick{v_j} 
    & \gate{2^0\theta_s} 
    & \gate{2^1\theta_s} 
    & \qw    & \qw
    & \gate{2^{n_E-2}\theta_s}
    & \gate{-2^{n_E-1}\theta_s}
    & \qw
    & \push{\rule{1em}{0em}}
    & \gate{-2^1\theta_s}
    & \gate{-2^2\theta_s}
    & \qw    
    & \qw
    & \gate{-2^{n_E-1}\theta_s}
    & \gate{2^{n_E}\theta_s}
    & \qw \\
\lstick{e_0} 
    & \ctrl{-1} 
    & \qw 
    & \qw
    & \qw
    & \qw    & \qw      & \qw 
    & \push{\rule{1em}{0em}}
    & \ctrl{0}
    & \qw
    & \qw
    & \qw
    & \qw    & \qw
    & \qw \\
\lstick{e_1} 
    & \qw 
    & \ctrl{-2} 
    & \qw
    & \qw
    & \qw    & \qw     & \qw 
    & \push{\rule{1em}{0em}}
    & \qw
    & \ctrl{0}
    & \qw
    & \qw
    & \qw
    & \qw    & \qw
    & \\
\lstick{\vdots}
    & & & & \lstick{\ddots}
    & & & & & & & & \lstick{\ddots}
    \\
\lstick{e_{n_E-3}} 
    & & & & & & & & & & & & & \ctrl{0} & \qw & \qw
    \\
\lstick{e_{n_E-2}} 
    & & & & & \ctrl{-5} & \qw & \qw & & & & & & \qw
    & \ctrl{0} &\qw
    \\
\lstick{e_{n_E-1}} 
    & \qw 
    & \qw 
    & \qw    & \qw     & \qw 
    & \ctrl{-6}
    & \qw
    & \push{\rule{1em}{0em}} 
    & \ctrl{-6}
    & \ctrl{-6}
    & \qw    & \qw
    & \ctrl{-6}
    & \ctrl{-6}
    & \qw
}
\]
\caption{Field-dependent acceleration block \(eW_j\). The phase labels indicate the elementary controlled-\(R_z\) angles arising from the two's-complement decomposition of the computed field register \(E_{\mathrm{fp}}\). The left portion implements the signed field contribution, while the portion after the
gap applies the additional \(|E|\)-dependent correction required by the \(A_1\) term in Eq.~\eqref{eq:A1_A2}. For the \(A_1\) block the scale is \(\theta_s=\theta_v/(RS)\); for the \(A_2\) block only the signed left portion is used, with \(\theta_s=\theta_v/S\).
}
\label{fig:eWj-block}
\end{figure}

A corresponding field-dependent acceleration block \(eW_j\) is shown in Fig.~\ref{fig:eWj-block}. This is the direct analogue of the velocity-dependent free-streaming block \(vW_j\) in Fig.~\ref{fig:vWj-block}, with the controlled phases now driven by the bits of the computed field register \(E_{\mathrm{fp}}\). 

Similarly, the field-controlled acceleration blocks are implemented in direct analogy with the velocity-controlled blocks in Figs.~\ref{fig:vV1_vV2}--\ref{fig:block_1}. After the field register has been prepared, \(vW_j\) and \(vU\) are replaced by their field-controlled counterparts \(eW_j\) and \(eU\), acting on the velocity register \(V\) with rotation controls supplied by the bits of \(E_{\mathrm{fp}}\). The \(A_1\) block \(eV_1\) uses the scale \(\theta_v/(RS)\), while the \(A_2\) block \(eV_2\) uses the scale \(\theta_v/S\), where \(\theta_v=\Delta t/\Delta v\).

Thus the full Block~2 update has the compute--apply--uncompute form
\begin{equation}
U_v(\Delta t) = U_E^\dagger\, U_v^{(E)}(\Delta t)\, U_E ,
\end{equation}
where \(U_E\) is the coherent field-evaluation circuit and \(U_v^{(E)}(\Delta t)\) is assembled from binary-controlled powers of \(eV_1\), followed by a single \(eV_2\), exactly as in Block~1. 

For multiple time steps, the one-step Strang operator in Eq.~\eqref{eq:strang_quantum} is applied repeatedly. Since adjacent half steps in \(U_x\) combine into a full step, the \(N_t\)-step evolution can be written as
\begin{equation}
U(N_t\Delta t) \approx
U_x\!\left(\frac{\Delta t}{2}\right)
\bigg[
U_v(\Delta t)\,
U_x(\Delta t)
\bigg]^{N_t-1}
U_v(\Delta t)\,
U_x\!\left(\frac{\Delta t}{2}\right),
\label{eq:strang_multistep}
\end{equation}
where
\begin{equation}
U_x(\Delta t) \equiv U_x\!\left(\frac{\Delta t}{2}\right)U_x\!\left(\frac{\Delta t}{2}\right).
\end{equation}

Let \(D_x\) and \(D_v\) denote the circuit depths of \(U_x\) and \(U_v\), respectively. Since \(U_x(\Delta t/2)\) and \(U_x(\Delta t)\) have identical circuit depth, the depth per time step is
\begin{equation}
\frac{D_{N_t}}{N_t} = D_v + \left(1+\frac{1}{N_t}\right)D_x .
\end{equation}
In particular,
\begin{equation}
\frac{D_{N_t}}{N_t} \longrightarrow D_v+D_x
\qquad \text{as} \qquad N_t\to\infty .
\end{equation}
Thus the two half steps contribute only an \(O(D_x/N_t)\) overhead per time step.

\subsection{Measurement and readout}
\label{subsec:measurement_readout}

After applying the Schr\"odingerised evolution (Eq.~\eqref{eq:strang_multistep}) in the auxiliary Fourier basis, we apply the inverse of the initial Fourier-basis transformation to the auxiliary register \(P\) to return to the \(p\)-grid. The physical solution is recovered by measuring \(P\) and selecting an appropriate non-negative \(p\)-slice~\cite{Jin2024PRL,Hu2024}. For the original auxiliary profile used here, we select the central \(p=0\) slice.

Let the post-IQFT state at time \(T=N_t\Delta t\) be
\begin{equation}
\ket{\Psi(T)} = \sum_{j,k,l} \psi_{jkl}(T) \ket{j}_X\ket{k}_V\ket{l}_P ,
\end{equation}
where \(p_l\) denotes the auxiliary grid defined in Eq.~\eqref{eq:p_grid}. The probability of measuring \(P=l\) is
\begin{equation}
P_l = \sum_{j,k} |\psi_{jkl}(T)|^2 ,
\end{equation}
and the corresponding normalised physical state is
\begin{equation}
\ket{\psi_l(T)}_{XV} = \frac{1}{\sqrt{P_l}}
\sum_{j,k} \psi_{jkl}(T)\ket{j}_X\ket{k}_V .
\end{equation}

For \(p_l>0\), the warped representation gives
\begin{equation}
\Phi(T,x_j,v_k,p_l) = e^{-p_l}f(T,x_j,v_k).
\end{equation}
Thus the selected \(p_l\)-slice is proportional to the physical solution. Since the factor \(e^{-p_l}\) is independent of \(j\) and \(k\), it cancels after normalisation, giving
\begin{equation}
\ket{\psi_l(T)}_{XV} \approx \frac{1}{\|f(T)\|} \sum_{j,k} 
f(T,x_j,v_k)\ket{j}_X\ket{k}_V .
\end{equation}

Reconstructing every amplitude of the full \(N_xN_v\)-dimensional solution would require tomography and is not assumed to be efficient. Alternative smooth auxiliary profiles, including the Hermite-spliced and error-function profiles, are discussed in Appendix~\ref{app:fourier_space_loading}.

\subsubsection{Memory-efficient readout for classical circuit simulation}
\label{subsec:flagged_readout}

For classical circuit validation, the quantum circuits are simulated using Qiskit Aer~\cite{qiskit2024} with its matrix-product-state (MPS) simulation method. The MPS representation avoids storing the full statevector explicitly which enables emulation of large circuits, although its computational cost still depends on the bond dimensions generated during the evolution. In this study, the MPS simulator is configured with a maximum bond dimension of \(\chi_{\mathrm{cap}}=512\) and a truncation threshold of \(10^{-12}\), using two CPU threads. These settings are fixed for a consistent assessment of the classical emulation cost.

Directly constructing the reduced density matrix on the full \(X,V,P\)
subsystem requires \(4^{n_x+n_v+n_p}\) complex entries. To reduce this memory cost, we append a clean flag qubit \(F\) after the Schr\"odingerised evolution and reversibly set
\[
F=1 \quad\Longleftrightarrow\quad P=q_{\mathrm{rec}},
\]
where \(q_{\mathrm{rec}}=N_p/2\) corresponds to \(p=0\) for the original
auxiliary profile.

The simulator then constructs only the reduced density matrix
\begin{equation}
\rho_{XVF} = \operatorname{Tr}_{P,W}\left(\ket{\Psi_F}\!\bra{\Psi_F}\right),
\end{equation}
where \(W\) denotes all field-arithmetic and workspace registers. The unnormalised postselected state is obtained from the flag-\(1\) block,
\begin{equation}
{\rho}_{XV}^{(q_{\mathrm{rec}})} = {}_F\!\bra{1}\rho_{XVF}\ket{1}_F,
\end{equation}
and its trace gives the postselection probability. After normalisation, the principal eigenvector is used as the recovered \(X,V\) amplitude vector when the conditional state is numerically pure.

This reduces the dense readout storage to \(4^{n_x+n_v+1}\), a reduction factor of \(4^{n_p-1}\). On quantum hardware, the \(P\)-register may of course be measured directly with the \(X,V\) outcomes postselected on \(q_{\mathrm{rec}}\).

\section{High-dimensional test case: 3D3V Vlasov equation}
\label{sec:vlasov_3d3v}

To test whether the explicit circuit architecture remains viable beyond the 1D1V Vlasov equation, we construct a low-resolution, three-dimensional in space and three-dimensional in velocity (3D3V),
\begin{equation}
    f=f(x,y,z,v_x,v_y,v_z,t),
\end{equation}
defined on a six-dimensional phase space. A direct tensor product grid with $N$ points in each coordinate contains \(N^6\) phase-space values, whereas the corresponding amplitude encoding requires \(6\log_2N\) physical qubits, in addition to the auxiliary and arithmetic registers.

\subsection{Anharmonic Penning trap}
\label{sec:penning_model}

A Penning trap confines charged particles by combining a static magnetic field with an electrostatic trapping potential. In the idealised configuration, a uniform magnetic field provides transverse magnetic confinement, while an electrostatic quadrupole potential provides confinement along the axial direction. 

We consider the prescribed-field kinetic equation
\begin{equation}
\begin{split}
\partial_t f &+ v_x \partial_x f + v_y \partial_y f + v_z \partial_z f \\
&+ \left(E_x +\omega_c v_y\right)\partial_{v_x}f
+ \left(E_y -\omega_c v_x\right)\partial_{v_y}f
+ E_z\partial_{v_z}f =0,
\end{split}
\label{eq:penning_3d3v}
\end{equation}
where \(\omega_c\) is a prescribed cyclotron-frequency parameter associated with a uniform magnetic field in the \(z\) direction. The electrostatic field is obtained from the quadrupole--quartic potential
\begin{equation}
    V_{\rm trap}(x,y,z) = A_2 \left(z^2-\frac{x^2+y^2}{2}\right) +A_4\left[z^4-3z^2(x^2+y^2)+\frac{3}{8}(x^2+y^2)^2\right].
\label{eq:penning_potential}
\end{equation}
The corresponding field components are
\begin{align}
E_x &= A_2x + A_4 \left( 6xz^2 -\frac{3}{2}x^3 -\frac{3}{2}xy^2\right),
\label{eq:penning_ex}\\
E_y &= A_2y + A_4 \left( 6yz^2 -\frac{3}{2}y^3 -\frac{3}{2}x^2y \right),
\label{eq:penning_ey}\\
E_z &= -2A_2z + A_4 \biggl( 6x^2z + 6y^2z - 4z^3 \biggr).
\label{eq:penning_ez}
\end{align}
Note that the mixed monomials \{\(xz^2,\, xy^2,\, yz^2,\, x^2y,\, x^2z,\, y^2z\)\} couple the three spatial coordinates, so the 3D3V problem cannot be factorised into three independent 1D1V problems.

\subsubsection{Circuit compilation of the coupled electric field}
\label{sec:boolean_penning_field}

In the preceding 1D1V Vlasov setup~(Sec.~\ref{sec:block2}), the cubic field was evaluated by first constructing signed coordinate and polynomial work registers. For the present 3D3V Vlasov proof of concept, we compile the known polynomial field directly from the binary digits of the spatial registers.

Let \(b_r^{(\mu)}\in\{0,1\}\) denote bit \(r\) of the register representing the spatial coordinate \(\mu\in\{x,y,z\}\). On a uniform cell-centred grid, the coordinate associated with a computational-basis state is
\begin{equation}
\mu = -\frac{L_\mu}{2} + \frac{\Delta\mu}{2} 
+ \Delta\mu\sum_{r=0}^{n_\mu-1}2^r b_r^{(\mu)}.
\label{eq:binary_cell_coordinate}
\end{equation}

Substituting Eq.~\eqref{eq:binary_cell_coordinate} into the analytic
expressions for \(E_x\), \(E_y\), and \(E_z\) (Eqs.~\eqref{eq:penning_ex}--\eqref{eq:penning_ez}) converts the spatial
polynomials into polynomials of the register bits. For example, the term \(xz^2\) becomes a sum of constant, linear, quadratic, and
cubic products involving bits from the \(X\) and \(Z\) registers. Since
the bits are in \(\{0,1\}\), repeated powers satisfy
\((b_r^{(\mu)})^m=b_r^{(\mu)}\) on computational basis. Hence, a
cubic spatial field produces Boolean monomials of degree at most three.

Collecting the bits of the three spatial registers into a single sequence \(b_1,\ldots,b_{n_s}\), where
\(n_s=n_x+n_y+n_z\), each field component can therefore be written as
\begin{equation}
E_\mu = c_{\mu,0} + \sum_i c_{\mu,i} b_i + \sum_{i<j} c_{\mu,ij} b_i b_j
+ \sum_{i<j<k} c_{\mu,ijk} b_i b_j b_k,
\qquad \mu\in\{x,y,z\}.
\label{eq:boolean_penning_field}
\end{equation}
The coefficients in Eq.~\eqref{eq:boolean_penning_field} are calculated only once during circuit compilation and no classical field evaluation is performed during the quantum time evolution.

The coefficients \(c\) are represented in signed fixed-point (fp) form with \(s\) fractional bits,
\begin{equation}
c^{\mathrm{fp}} = \operatorname{round}\!\left(2^s c\right).
\label{eq:boolean_penning_coeff_fp}
\end{equation}
For each field component, these fixed-point contributions are accumulated into a signed field register \(E\).

To implement a linear term \(c_i b_i\), the spatial qubit \(b_i\) is used directly as a control. When \(b_i=1\), a temporary register \(W\) is prepared in the computational basis representing the fixed-point coefficient \(c_i^{\mathrm{fp}}\). The register \(W\) is then reversibly added into the field register \(E\) and subsequently returned to \(|0\rangle\). For a quadratic or cubic term, the product of the required spatial bits is first computed into a single reusable Boolean ancilla. Schematically,
\begin{equation}
\text{compute monomial}
\;\longrightarrow\;
\text{add fixed-point coefficient to }E
\;\longrightarrow\;
\text{uncompute monomial}.
\label{eq:boolean_penning_compute_add}
\end{equation}
In this way, the mixed-coordinate terms of the Penning field can be represented without storing a separate numerical register for \(xz^2\), \(xy^2\), or other coordinate products.

Finally, the field register contains the fixed-point value \(E_\mu^{\mathrm{fp}}(X,Y,Z)\). Then, the three electric components are applied sequentially via
\begin{equation}
U_{E_x}^{\dagger} U_{v_x}^{(E_x)} U_{E_x},
\qquad
U_{E_y}^{\dagger} U_{v_y}^{(E_y)} U_{E_y},
\qquad
U_{E_z}^{\dagger} U_{v_z}^{(E_z)} U_{E_z},
\label{eq:penning_field_compute_apply}
\end{equation}
where \(U_{E_\mu}\) computes the corresponding field component and
\(U_{v_\mu}^{(E_\mu)}\) denotes the field-controlled velocity update. The field components \(E_x\), \(E_y\), and \(E_z\) are computed, applied to the corresponding velocity update, and uncomputed sequentially, allowing the same field and workspace registers to be reused for all three components.

\subsubsection{Low-resolution 3D3V circuit}
\label{sec:penning_circuit_instance}

For the proof of concept, we allocate two qubits to each physical coordinate,
\begin{equation}
    n_x=n_y=n_z=n_{v_x}=n_{v_y}=n_{v_z}=2,
\label{eq:penning_poc_resolution}
\end{equation}
corresponding to four grid points per coordinate and \(N_{\rm grid}=4^6=4096\) points in the discretised six-dimensional phase space. The spatial domains have lengths \(L_x=L_y=L_z=2\), while the velocity domains have lengths \(L_{v_x}=L_{v_y}=L_{v_z}=4\). The prescribed field parameters are \(A_2=0.45\), \(A_4=0.025\), and \(\omega_c=2\).

For this low-resolution case with \(s=7\) fixed-point precision, the binary field construction produces 20 non-constant terms for each of \(E_x\), \(E_y\), and \(E_z\). These are linear, quadratic, or cubic products of the spatial register bits. A second 10-qubit register \(W\) is used to prepare the fixed-point constants required by the reversible additions. Since the three field components are computed and uncomputed sequentially, this workspace is shared between
\(E_x\), \(E_y\), and \(E_z\).

For baseline calculations, we allocate \(n_p=5\) qubits to the auxiliary register with \(R=6\), giving \(2^{n_p}=32\) points in
the auxiliary coordinate. Altogether, the evolution circuit
contains 39 algorithmic logical qubits, of which 12 encode the physical 3D3V phase space.


The initial condition is chosen as a product of 1D Gaussian distributions in the six phase space coordinates, 
\begin{equation}
    f_0(x,y,z,v_x,v_y,v_z) = \mathcal{Z} \prod_{\mu\in\{x,y,z,v_x,v_y,v_z\}} \exp\left[ -\frac{(\mu-\mu_0)^2}{2\sigma_\mu^2} \right],
\label{eq:penning_initial}
\end{equation}
where \(\mathcal{Z}\) is the discrete normalisation constant. The centres and widths are, in \((x,y,z,v_x,v_y,v_z)\) order,
\begin{equation}
    \boldsymbol{\mu}_0 = (0.20,0,0.10,0,0.40,0), \qquad \boldsymbol{\sigma} = (0.35,0.35,0.40,0.90,0.90,0.90).
\label{eq:penning_initial_parameters}
\end{equation}
Because the initial distribution is a product of six 1D Gaussians, each register can be initialised separately, thus avoiding the need to amplitude-load a full 6D oracle.

The 3D3V evolution contains three types of transport. The spatial streaming terms \(v_x\partial_x\), \(v_y\partial_y\), and \(v_z\partial_z\) are implemented using the velocity-controlled construction of Sec.~\ref{sec:block1}. The electric-field terms are implemented sequentially using the compute--apply--uncompute procedure of Eq.~\eqref{eq:penning_field_compute_apply}.

The magnetic contribution couples the two transverse velocity coordinates,
\begin{equation}
    \mathcal{L}_B f = \omega_c v_x\partial_{v_y}f-\omega_c v_y\partial_{v_x}f.
\label{eq:penning_magnetic}
\end{equation}
We implement this contribution using the symmetric splitting
\begin{equation}
    U_B(\tau) \approx U_{v_x\leftarrow v_y}\left(\frac{\tau}{2}\right) U_{v_y\leftarrow v_x}(\tau) U_{v_x\leftarrow v_y}\left(\frac{\tau}{2}\right),
\label{eq:penning_magnetic_split}
\end{equation}
where the arrow indicates which velocity coordinate is updated and which coordinate supplies the coefficient. The outer updates are proportional to \(+\omega_c v_y\), while the central update is proportional to \(-\omega_c v_x\).

Accordingly, the Strang-split time step is
\begin{equation}
    U_{\mathrm{3D3V}}(\Delta t) \approx U_K\left(\frac{\Delta t}{2}\right) U_B\left(\frac{\Delta t}{2}\right) U_E(\Delta t) U_B\left(\frac{\Delta t}{2}\right) U_K\left(\frac{\Delta t}{2}\right),
\label{eq:penning_strang}
\end{equation}
where \(U_K\), \(U_E\), and \(U_B\) denote the configuration-space
streaming, electric-field, and magnetic updates, respectively.

\section{Results}
\label{sec:results_1d1v}

We evaluate both logical circuit resources and the practical limits of matrix-product-state (MPS) emulation. While qubit counts, circuit depth, and gate counts characterise the quantum implementation, MPS runtime and bond dimension also depend on the entanglement generated during evolution.

\subsection{1D1V prescribed-field circuit validation}
\label{sec:results_1d1v_validation}

We first validate the complete 1D1V Schr\"odingerised circuit developed in Sec.~\ref{sec:1d1v_theory} against classical computation. The spatial and velocity registers each contain five qubits, \(n_x=n_v=5\), corresponding to \(N_x=N_v=32\) grid points, while the auxiliary Schr\"odingerisation register contains \(n_p=7\) qubits. We take \(L_x=L_v=6\), \(R=6\), and use the original auxiliary profile \(f_p(p)=e^{-|p|}\), for which the physical solution is recovered from the central \(p=0\) slice (Sec.~\ref{subsec:measurement_readout}). The initial distribution is sampled on the cell-centred grid, while the prescribed field is evaluated using the centred-integer coordinates defined in Sec.~\ref{sec:field_register}. The field coefficients are represented using the same signed fixed-point convention in both the quantum circuit and the classical reference.

For this benchmark, we set \(a_0=\kappa=1\) and use fixed-point precision \(s=6\), giving \(\alpha_{\rm fp}=12\) and \(\beta_{\rm fp}=0\). Thus, coefficient rounding suppresses the cubic contribution in this particular instance, although the circuit retains the general cubic-field arithmetic. The reduced test in Appendix~\ref{app:grid_sensitivity} additionally
validates a non-zero cubic coefficient.

The initial phase-space distribution is chosen as
\begin{equation}
f_0(x,v) = \mathcal{Z}^{-1}\exp \left[-\beta_0 V_{\rm dw}(x)\right]\exp\left(-\frac{v^2}{2\sigma^2}\right),
\label{eq:results_initial_condition}
\end{equation}
where \(\mathcal{Z}\) denotes the discrete normalisation constant, with \(\beta_0=2\) and \(\sigma=1\). Together with \(L_x=L_v=6\), the initial distribution is well within the truncated computational domain. 

\begin{figure}[htbp]
\centering
\includegraphics[width=1\linewidth]{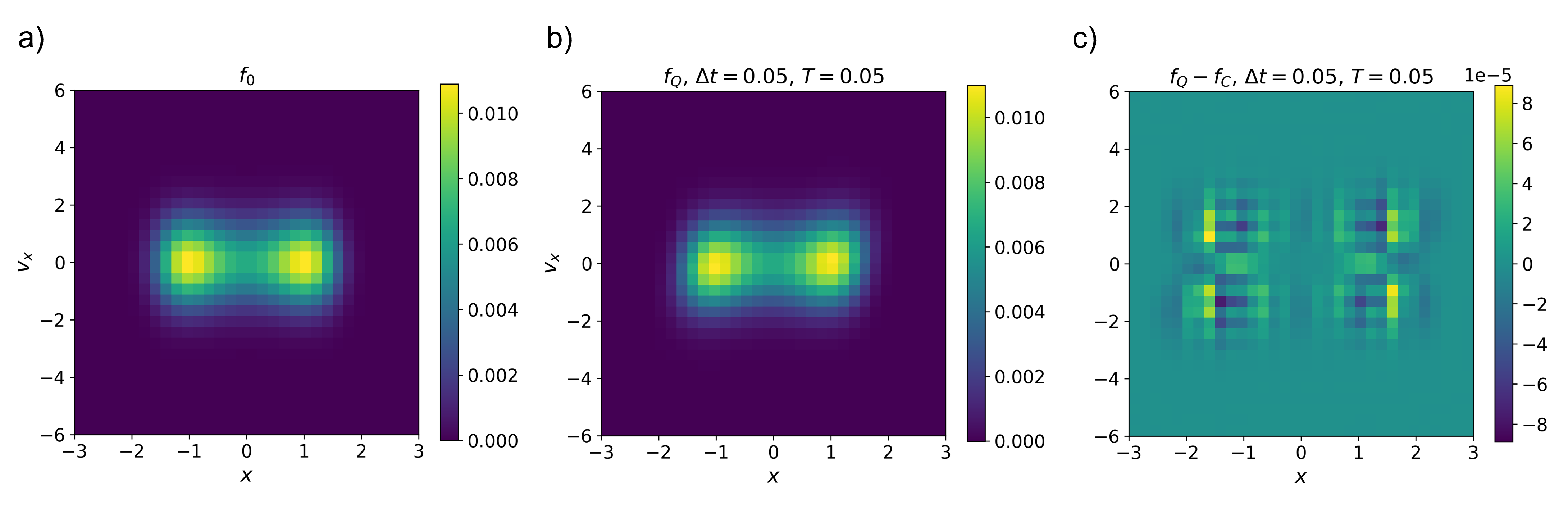}
\caption{ Phase-space validation of the 1D1V Schr\"odingerised Vlasov circuit for \(n_x=n_v=5\), \(n_p=7\), \(N_t=1\), \(\Delta t=0.05\), and \(T=0.05\). (a) Initial distribution \(f_0(x,v_x)\). (b) Recovered quantum distribution \(f_Q(x,v_x,T)\). (c) Pointwise difference between the quantum and scheme-matched classical solutions, \(f_Q-f_C\). The difference is of order \(10^{-5}\), compared with distribution amplitudes of order \(10^{-2}\). }
\label{fig:1d1v_phase_space}
\end{figure}

Fig.~\ref{fig:1d1v_phase_space} compares the initial distribution with the quantum result after one Strang step using \(N_t=1\), \(\Delta t=0.05\), and \(T=0.05\). The recovered quantum distribution retains the two-lobe phase-space structure of the initial state slightly shifted by the free-streaming and field-acceleration blocks.

Agreement with the classical reference is shown directly in Fig.~\ref{fig:1d1v_phase_space}(c). The pointwise difference \((f_Q-f_C)\sim \mathcal O(10^{-5})\) is concentrated at hotspots and less than phase-space densities of order \(\mathcal O(10^{-2})\).  For this run, the relative \(L_2\) discrepancy between the recovered quantum solution and the matching classical solution is \(\epsilon_{Q-C}=4.9\times10^{-3}\). The residual difference contains the accumulated effects of the Schr\"odingerised circuit approximation, finite auxiliary register, fixed-point coefficient representation, and classical tensor-network emulation.

Next, we project onto the spatial and velocity coordinates and define the respective discrete marginals
\begin{align}
\rho(x_j,t) &= \sum_k f(x_j,v_k,t),\\
g(v_k,t) &= \sum_j f(x_j,v_k,t).
\label{eq:1d1v_marginals}    
\end{align}

\begin{figure}[htbp]
\centering
\includegraphics[width=0.9\linewidth]{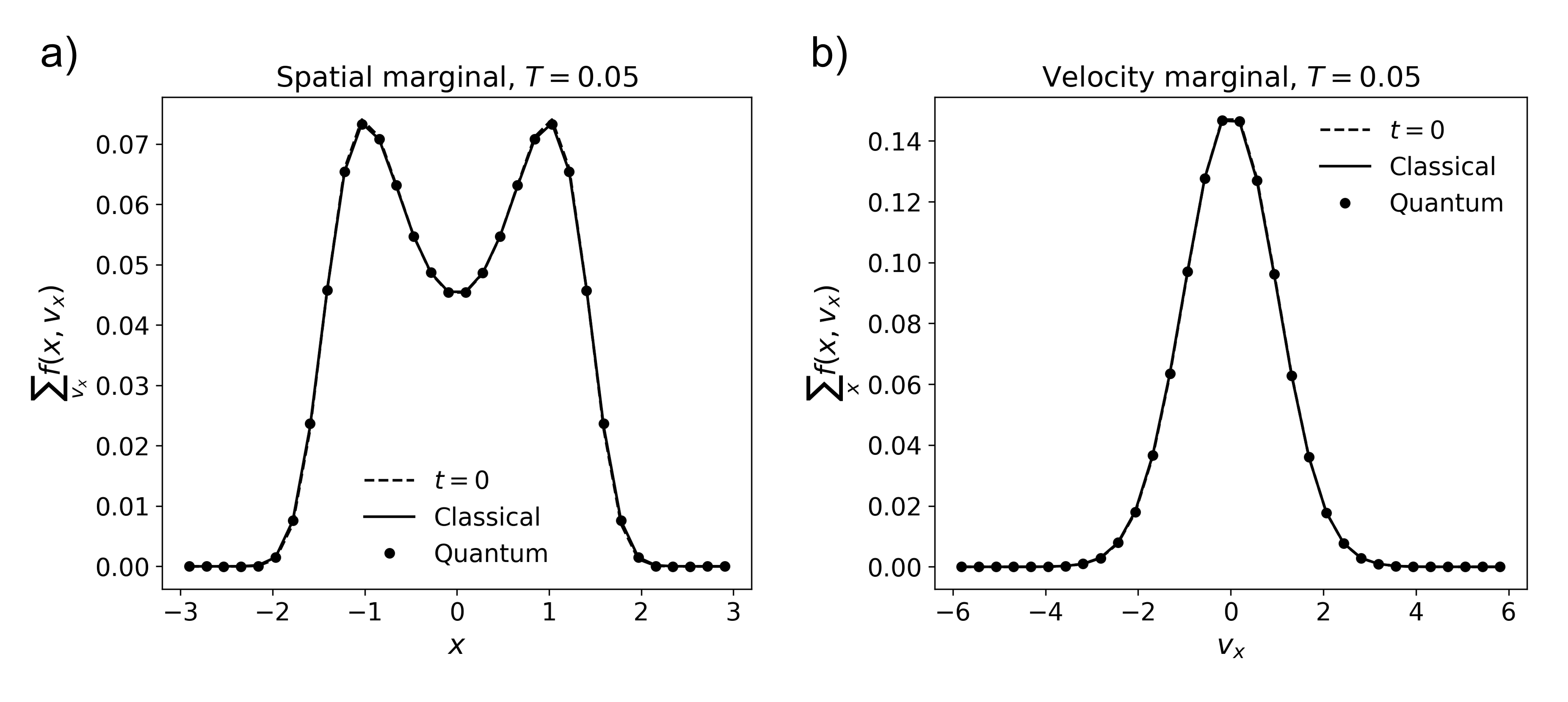}
\caption{ Marginal distributions for the 1D1V validation case at \(T=0.05\). (a) Spatial marginal \(\rho(x)=\sum_{v_x}f(x,v_x)\). (b) Velocity marginal \(g(v_x)=\sum_x f(x,v_x)\). The initial distribution (\(t=0\), dashed), scheme-matched classical solution (solid), and recovered quantum solution (markers) are shown. }
\label{fig:1d1v_marginals}
\end{figure}

Fig.~\ref{fig:1d1v_marginals} compares these quantities for the initial distribution, the classical solution, and the quantum result at
\(T=0.05\). The spatial marginal preserves the
two maxima associated with the two localised populations, while the
velocity marginal remains approximately symmetric about \(v=0\).  The
near overlap of the \(t=0\) and \(T=0.05\) curves is consistent with the
short simulated time and confirms that the small quantum--classical
difference in Fig.~\ref{fig:1d1v_phase_space}(c) is not produced by a
qualitatively incorrect redistribution of the phase-space density.

Together, Figs.~\ref{fig:1d1v_phase_space} and \ref{fig:1d1v_marginals} provide an end-to-end validation of the Schr\"odingerisation workflow, following amplitude preparation, auxiliary Fourier transformation, velocity-dependent free streaming, coherently field-controlled acceleration, uncomputation of the arithmetic workspace, and recovery of the physical \(p\)-slice.

\subsubsection{Resource scaling and temporal behaviour}
\label{sec:results_1d1v_resources}

\begin{figure}[htbp]
\centering
\includegraphics[width=1\linewidth]{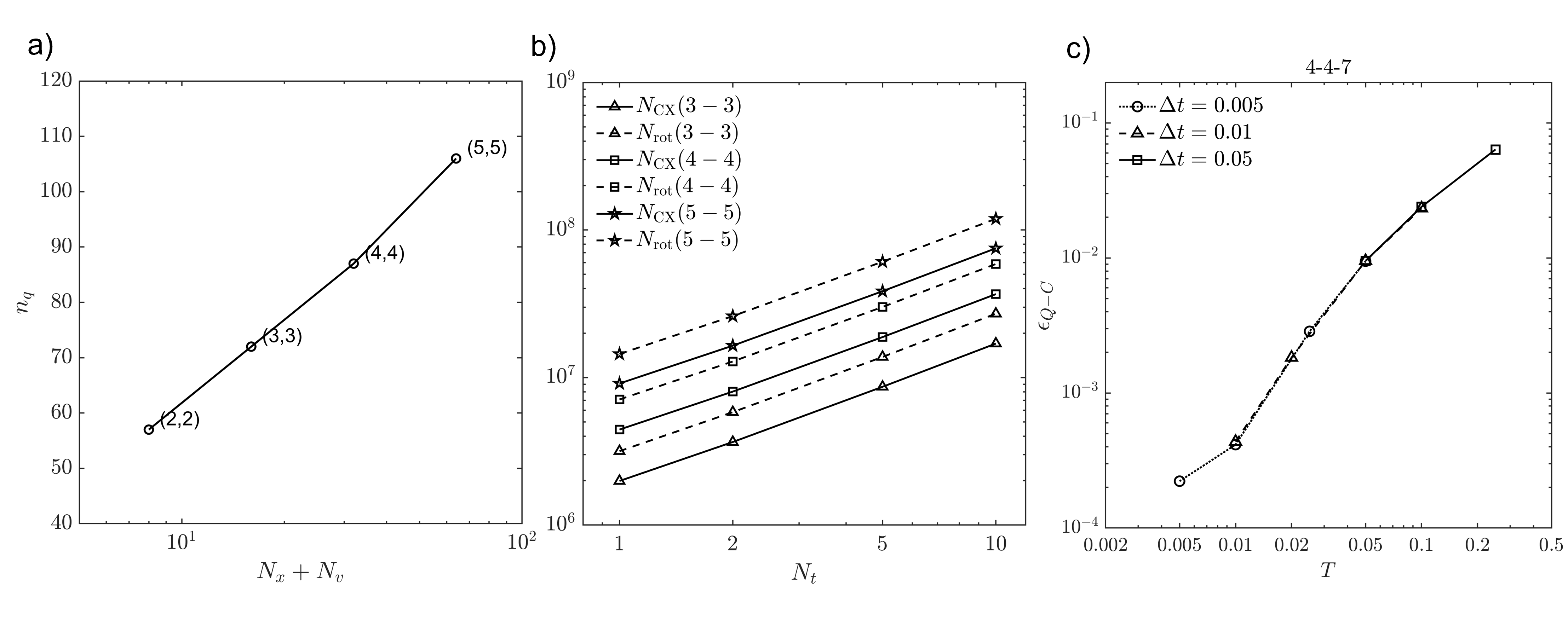}
\caption{
Resource scaling and temporal behaviour of the 1D1V Schr\"odingerised
Vlasov circuit.
(a) Peak logical-qubit count \(n_q\) as the spatial and velocity grid
resolutions are increased together from \(n_x=n_v=2\) to \(5\), with
\(n_p=7\) fixed.  The horizontal axis shows the corresponding combined
number of physical grid points, \(N_x+N_v\).
(b) Decomposed CNOT count \(N_{\mathrm{CX}}\) and rotation-like gate count
\(N_{\mathrm{rot}}\) as functions of the number of Strang steps \(N_t\) for
\((n_x,n_v)=(3,3)\), \((4,4)\), and \((5,5)\), with \(n_p=7\).
Both resource measures grow approximately linearly with \(N_t\).
(c) Relative quantum--classical \(L_2\) discrepancy
\(\epsilon_{Q-C}\) versus total simulated time \(T=N_t\Delta t\) for the
\((n_x,n_v,n_p)=(4,4,7)\) circuit using
\(\Delta t=0.005\), \(0.01\), and \(0.05\).
Results obtained with different time step sizes are close at common values
of \(T\), while the discrepancy increases with accumulated evolution time.
}
\label{fig:1d1v_resources}
\end{figure}

Here, we estimate the circuit resources required by the explicit 1D1V
construction and the behaviour of the quantum--classical discrepancy. Fig.~\ref{fig:1d1v_resources}(a) shows the peak number of logical qubits required by the complete circuit as the spatial and velocity resolutions are increased together, \(n_x=n_v=2,\ldots,5\), with the auxiliary register size fixed at \(n_p=7\).  The corresponding physical grids range from \(N_x=N_v=4\) to \(N_x=N_v=32\).  The total circuit width increases from approximately \(57\) to \(106\) qubits over this range. For fixed polynomial degree and auxiliary precision, the total qubit requirement grows approximately linearly with \(n_x+n_v\), and hence only logarithmically with the number of phase-space grid points in each coordinate (Lemma~\ref{lem:qubit_scaling} and Remark~\ref{rem:qubit_scaling}).

The gate-count depends on the number of time steps (Fig.~\ref{fig:1d1v_resources}(b)). We report the number of CNOT gates, \(N_{\mathrm{CX}}\), together with the rotation-like gate count \(N_{\mathrm{rot}} = N_u + N_{\mathrm{phase}} + N_{R_z} + N_{CP}\) obtained after decomposing the composite circuit blocks, where \(u\), \(\mathrm{phase}\), \(R_z\), and \(CP\) denote the corresponding parameterised Qiskit gates (\(N_{\mathrm{phase}}\) counts the number of \(p\) gates). The quantity \(N_{\mathrm{rot}}\) is intended as a proxy for non-Clifford workload in a fault-tolerant implementation and should therefore be interpreted as logical circuit resources.

For each of the \((n_x,n_v)\) cases, we see that gate counts increase approximately linearly with \(N_t\). This agrees with Eq.~\eqref{eq:strang_multistep}, as each additional time step adds one free-streaming block~(Sec.~\ref{sec:block1}) and one field-driven block~(Sec.~\ref{sec:block2}), excluding the two half steps at the beginning and end of the evolution. Increasing the phase-space resolution only shifts the curves upward because the velocity-dependent transport and the field subroutines acting on the ancilla registers do not scale with the number of time steps.

Finally, Fig.~\ref{fig:1d1v_resources}(c) shows the relative
quantum--classical discrepancy
\begin{equation}
    \epsilon_{Q-C} = \left\| \frac{f_Q}{\|f_Q\|_2} -
    \frac{f_C}{\|f_C\|_2} \right\|_2.
\label{eq:Q-C_discrepancy}
\end{equation}
for the \((n_x,n_v,n_p)=(4,4,7)\) circuit as a function of the total
simulated time \(T=N_t\Delta t\). The discrepancy increases monotonically with \(T\), from \(10^{-4}\) to several \(10^{-2}\) by \(T=0.25\). Within the range tested, the total quantum--classical discrepancy depends more strongly on the total evolution time itself than on the particular Strang decomposition into multiple time steps.

Note that \(\epsilon_{Q-C}\) measures agreement between the explicit Schr\"odingerised circuit and its classical counterpart, rather than the total discretisation error with respect to the continuum Vlasov equation.

\subsection{3D3V proof of concept validation}
\label{sec:results_3d3v}

In this section, we validate the low-resolution 3D3V circuit instance defined in Sec.~\ref{sec:penning_circuit_instance} against classical calculation. Since the full 3D3V solution depends on six phase-space coordinates, we visualise the result through two-dimensional pair marginals obtained by summing over the remaining four coordinates. We consider
\begin{align}
f_{zv_z}(z,v_z,t) &= \sum_{x,y,v_x,v_y} f(x,y,z,v_x,v_y,v_z,t), \\
f_{v_xv_y}(v_x,v_y,t) &= \sum_{x,y,z,v_z} f(x,y,z,v_x,v_y,v_z,t).
\end{align}
The first retains the axial phase-space coordinates \((z,v_z)\), for which the velocity evolution is driven by the electric field, while the second retains the two transverse velocity coordinates \((v_x,v_y)\) which are related to transverse magnetic coupling.

\begin{figure}[htbp]
    \centering
    \includegraphics[width=\linewidth]{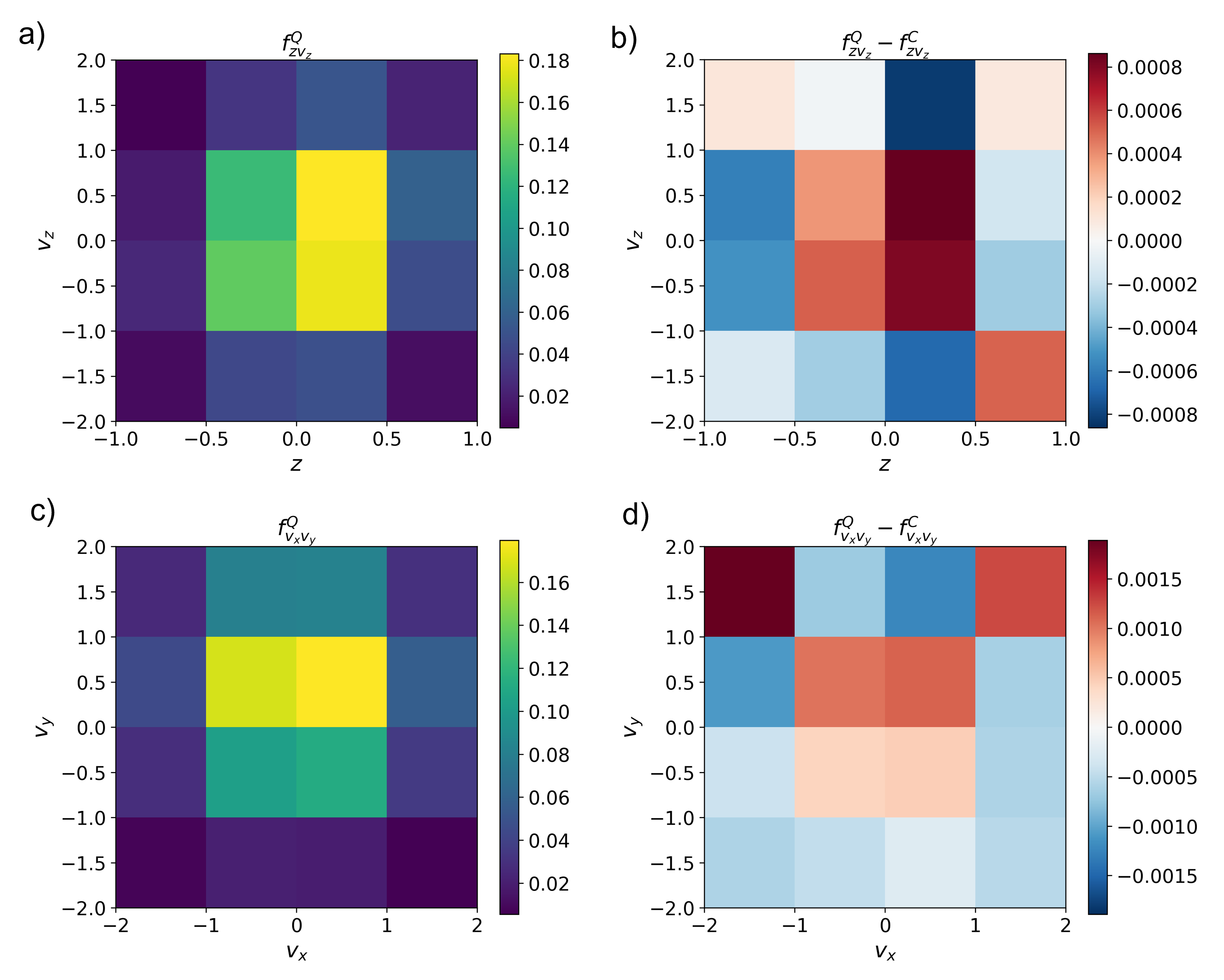}
    \caption{Representative pair marginals for the 3D3V proof of concept calculation with \(n_p=5\), \(N_t=2\), \(\Delta t=0.05\), and \(T=0.1\).
    (a) Quantum axial marginal \(f^Q_{zv_z}\) and
    (b) its pointwise difference from classical result, \(f^Q_{zv_z}-f^C_{zv_z}\).
    (c) Quantum transverse-velocity marginal \(f^Q_{v_xv_y}\) and
    (d) the corresponding difference.}
    \label{fig:3d3v_pair_marginals}
\end{figure}

Fig.~\ref{fig:3d3v_pair_marginals} shows these marginals for \(n_p=5\), \(N_t=2\), \(\Delta t=0.05\), and \(T=0.1\). In both
cases, the quantum result qualitatively reproduces the classical solution, with residual differences small relative to the marginal amplitudes.

Table~\ref{tab:3d3v_validation} summarizes the gate-level simulation cases performed with an allocated memory of 440~GB per job and a walltime limit of 120~hours. The baseline cases (Eq.~\eqref{eq:penning_poc_resolution}) mainly use \(n_p=5\), including one \(n_p=7\) case to examine the effect of increasing the size of the auxiliary register on solution error \(\epsilon_{Q-C}\) (Eq.~\eqref{eq:Q-C_discrepancy}). The table also reports the final maximum MPS bond dimension and Aer simulation time to indicate the cost of classically verifying the complete quantum circuit.

\begin{table}[t]
\centering
\caption{Simulations of low-resolution 3D3V circuits (Sec.~\ref{sec:penning_circuit_instance}). The relative mass error is evaluated with respect to the classical solution. \(\chi_{\max}\) is the maximum bond dimension of the final MPS representation. Reported MPS time is the Aer simulation time of classical tensor networks.}
\label{tab:3d3v_validation}
\begin{tabular}{cccccccc}
\hline
\(n_p\) &
\(N_t\) &
\(\Delta t\) &
\(T\) &
\(\epsilon_{Q-C}\) &
mass error &
\(\chi_{\max}\) &
MPS time (h)
\\
\hline
5 & 1 & 0.005 & 0.005 &
\(9.78\times10^{-4}\) &
\(1.14\times10^{-5}\) &
61 & 6.9
\\
5 & 2 & 0.005 & 0.010 &
\(1.69\times10^{-3}\) &
\(2.64\times10^{-5}\) &
85 & 16.8
\\
5 & 5 & 0.005 & 0.025 &
\(2.72\times10^{-3}\) &
\(5.76\times10^{-5}\) &
119 & 58.5
\\
5 & 1 & 0.050 & 0.050 &
\(7.74\times10^{-3}\) &
\(5.65\times10^{-5}\) &
159 & 22.5
\\
5 & 2 & 0.050 & 0.100 &
\(3.88\times10^{-2}\) &
\(2.68\times10^{-4}\) &
401 & 59.3
\\
7 & 2 & 0.005 & 0.010 &
\(1.78\times10^{-3}\) &
\(1.64\times10^{-4}\) &
207 & 97.0
\\
\hline
\end{tabular}
\end{table}

For the \(n_p=5\), \(\Delta t=0.005\) cases, the full six-dimensional discrepancy increases from \(9.78\times10^{-4}\) at \(T=0.005\) to \(1.69\times10^{-3}\) at \(T=0.01\) and \(2.72\times10^{-3}\) at \(T=0.025\). For the larger step \(\Delta t=0.05\), the discrepancy is \(7.74\times10^{-3}\) at \(T=0.05\) and \(3.88\times10^{-2}\) at \(T=0.1\). The relative mass error remains below \(3\times10^{-4}\) in all completed cases. These results show increasing quantum--classical discrepancy with evolution time, while the relative mass error remains below \(3\times10^{-4}\).

For \(n_p=5\) and \(N_t=1\), increasing \(\Delta t\) from \(0.005\) to \(0.05\) increases
the final maximum MPS bond dimension from \(61\) to \(159\), while the Aer simulation time increases from \(6.9\) to \(22.5\) hours. Although the logical circuit topology is
unchanged, the modified rotation angles lead to quantum states with different MPS representation costs. This shows that the computational difficulty of gate-level MPS emulation cannot be inferred solely from circuit depth or gate count.

We compare the auxiliary-register size at \(N_t=2\), \(\Delta t=0.005\), and \(T=0.01\). Increasing \(n_p\) from 5 to 7 changes the full six-dimensional discrepancy from \(1.69\times10^{-3}\) to \(1.78\times10^{-3}\), so no improvement is observed for this particular low-resolution instance. In contrast, the classical MPS cost increases substantially as the simulation time rises from approximately \(16.8\) h to \(97.0\) h and the maximum bond dimension \(\chi_{\max}\) increases from 85 to 207. 

The probability \(P_{\rm rec}\) of obtaining the selected \(p=0\) recovery slice remains between approximately \(0.759\) and \(0.827\) for the completed \(n_p=5\) cases. However, for the \(n_p=7\), \(N_t=2\) case, \(P_{\rm rec}\) decreases to approximately \(0.285\). Thus, increasing the auxiliary-register size also increases the postselection overhead at readout.

As an additional verification, the conditional fraction of the field workspace recovered in \(|0\rangle\) exceeds \(0.99999999996\) for every completed calculation. This confirms that the field and arithmetic work registers are cleanly uncomputed before readout.


The accessible parameter range is constrained by the computational cost of classical MPS emulation. Several more demanding circuit instances did not complete within the imposed 120-hour walltime limit. These observations characterise the practical verification limits under the specified simulation settings and computational resources, but not quantum hardware execution times.

\subsubsection{3D3V resource scaling}
\label{sec:3d3v_resources}

Table~\ref{tab:3d3v_resources} summarises the logical resources of the low-resolution 3D3V circuit. At fixed \(n_p=5\), the decomposed depth, CNOT count, and rotation-like gate count increase approximately linearly with \(N_t\), consistent with repeated application of the split evolution blocks.

The size of the auxiliary register has a more significant effect. At fixed \(N_t=2\), increasing \(n_p\) from 5 to 7 adds only two logical qubits, but increases the decomposed depth and both reported gate counts by approximately a factor of four. This scaling behaviour is consistent with the binary-controlled powers of the Blocks and that increasing the auxiliary resolution directly increases circuit depth (Fig.~\ref{fig:block_1}).

\begin{table}[htbp]
\centering
\caption{Logical circuit resources for the low-resolution 3D3V construction. The phase-space resolution is fixed at two qubits per coordinate.}
\label{tab:3d3v_resources}
\begin{tabular}{cccccc}
\hline
\(n_p\) & \(N_t\) & logical qubits & depth & \(N_{\rm CX}\) & \(N_{\rm rot}\) \\
\hline
5 & 1 & 39 & \(9.95\times10^{5}\) & \(5.77\times10^{5}\) & \(8.43\times10^{5}\) \\
5 & 2 & 39 & \(1.92\times10^{6}\) & \(1.10\times10^{6}\) & \(1.60\times10^{6}\) \\
5 & 5 & 39 & \(4.68\times10^{6}\) & \(2.66\times10^{6}\) & \(3.88\times10^{6}\) \\
7 & 2 & 41 & \(7.71\times10^{6}\) & \(4.35\times10^{6}\) & \(6.54\times10^{6}\) \\
\hline
\end{tabular}
\end{table}

\section{Discussion}
\label{sec:discussion}

We implemented an explicit Schr\"odingerised circuit for Vlasov transport with non-constant coefficients coherently. In the 1D1V problem, the
streaming coefficient is the encoded velocity \(v\), while the acceleration coefficient is the non-linear field \(E(x)\). These two cases require different circuit constructions. The affine dependence of \(v\) on the velocity-register bits can be compiled directly into controlled rotations, whereas the cubic field is evaluated coherently using reversible fixed-point arithmetic. The 3D3V construction extends this idea to an anharmonic Penning field problem that contains mixed spatial terms, while the magnetic contribution couples \(v_x\) and \(v_y\). This ensures the resulting evolution is not an otherwise trivial tensor product of independent 1D1V problems. 

A six-dimensional grid grows exponentially in the number of phase-space coordinates, while its amplitude encoding requires only the sum of the register widths. However, this does not by itself imply an efficient quantum algorithm. The transport operators must still be compiled into gates, and their cost depends on the algebraic structure of the coefficients. The present polynomial examples remain tractable because the velocity coefficient is affine and the electric fields have fixed polynomial degree. More general fields would require a correspondingly more expensive coherent representation.

The resource accounts identify the auxiliary Schrödingerisation register as a separate and more severe source of circuit cost. Increasing the number of time steps produces the expected approximately linear increase in gate count, since the split evolution blocks are repeated. In contrast, increasing \(n_p\) presents a far more significant cost on circuit depth, because the logarithmic encoding of the phase-space grid coexists with a circuit depth that is exponential in \(n_p\). Hence, the practical utility of the method depends on whether the required auxiliary resolution can remain sufficiently modest for the
problem and accuracy of interest.

The numerical results also show that the number of auxiliary phase qubits \(n_p\) should not be regarded as a convergence parameter in isolation. In a representative 3D3V instance, increasing \(n_p\) does not reduce the quantum--classical discrepancy, while the probability of post-selection decreases and the circuit cost increases substantially. This suggests that given sufficient auxiliary representation, increasing \(n_p\) alone need not improve the final solution.

The auxiliary phase profile study in Appendix~\ref{app:fourier_space_loading} also exposes a related trade-off. The optimal error-function profile has a nearly geometric positive tail, which has the potential to be loaded much more economically than a generic amplitude vector. However, the smoother profile removes the fixed central recovery point which is otherwise available for \(e^{-|p|}\), and this creates post-selection ambiguity at recovery. This means that auxiliary state preparation and state readout cannot be treated as independent design problems. 

A further limitation in the present construction is that the fields are prescribed. In a self-consistent Vlasov--Poisson or Vlasov--Maxwell system, the force field depends on the evolving distribution itself. Extending the present construction would therefore require a coherent field solve and feedback of that result into the transport circuit, or alternatively a hybrid quantum--classical update. 

Finally, the practical utility of amplitude encoding depends on what is required from the solution. The full \(N_xN_v\) or six-dimensional phase-space array is represented compactly in the quantum state, but recovering every amplitude would remove this advantage through tomography. The reality may be that only selected observables, moments, marginal probabilities, or other quantities can be estimated directly from the encoded state.

The present study demonstrates the explicit gate-level realisation of Schr\"odingerised Vlasov transport with prescribed, variable coefficients in both 1D1V and coupled 3D3V settings. Classical MPS emulation verifies selected low-resolution instances while revealing
substantial computational limitations as the evolution time and auxiliary-register size increase. Our results identify the auxiliary-register overhead as a key quantum resource challenge. Future work should focus on reducing this overhead, developing more efficient
coefficient compilation, and extracting physically relevant observables without reconstructing the full phase-space distribution.

\section*{Acknowledgements}
This project is supported by the National Research Foundation, Singapore through the National Quantum Office, hosted in A*STAR, under the Hybrid Quantum Classical Computing (HQCC) 1.0 Funding Initiative (S24Q7D7001) and the Advanced Quantum Algorithms and Solutions (AQAS) Funding Initiative (S25Q9DA001). The computational work for this article was partially performed on resources of the National Supercomputing Centre (NSCC), Singapore (\url{https://www.nscc.sg}).

\section*{Declaration of Competing Interest}
The authors declare that they have no known competing financial interests or personal relationships that could have appeared to influence the work reported in this paper.

\section*{Data Availability Statement}
The source code for the 1D1V simulations, including postprocessing scripts, is available at \url{https://github.com/Soracak/schrodingerised-vlasov}. 

\clearpage
\bibliographystyle{unsrt}
\bibliography{bib}

@article{Palmroth2025,
  author  = {Palmroth, Minna and Ganse, Urs and Pfau-Kempf, Yann
             and Battarbee, Markus and Alho, Markku and N{\"a}ttil{\"a}, Joonas
             and Zaitsev, Ivan and Cozzani, Giulia and Papadakis, Konstantinos
             and Kotipalo, Leo and Zhou, Hongyang and Turc, Lucile
             and Hoilijoki, Sanni and Grandin, Maxime
             and P{\"a}nk{\"a}l{\"a}inen, Lauri and Sandroos, Arto
             and von Alfthan, Sebastian},
  title   = {{V}lasov methods in space physics and astrophysics},
  journal = {Living Reviews in Computational Astrophysics},
  volume  = {11},
  number  = {1},
  pages   = {3},
  year    = {2025},
  doi     = {10.1007/s41115-025-00024-0}
}

@article{EinkemmerMoriggl2023,
  author  = {Einkemmer, Lukas and Moriggl, Alexander},
  title   = {Semi-{L}agrangian 4d, 5d, and 6d kinetic plasma simulation
             on large-scale {GPU}-equipped supercomputers},
  journal = {The International Journal of High Performance Computing Applications},
  volume  = {37},
  number  = {2},
  pages   = {180--196},
  year    = {2023},
  doi     = {10.1177/10943420221137599}
}

@article{Sato2024,
  author  = {Sato, Yuki and Kondo, Ruho and Hamamura, Ikko
             and Onodera, Tamiya and Yamamoto, Naoki},
  title   = {Hamiltonian simulation for hyperbolic partial differential
             equations by scalable quantum circuits},
  journal = {Physical Review Research},
  volume  = {6},
  number  = {3},
  pages   = {033246},
  year    = {2024},
  doi     = {10.1103/PhysRevResearch.6.033246}
}

@article{Hu2024,
  author  = {Hu, Junpeng and Jin, Shi and Liu, Nana and Zhang, Lei},
  title   = {Quantum {C}ircuits for partial differential equations via
             {S}chr{\"o}dingerisation},
  journal = {Quantum},
  volume  = {8},
  pages   = {1563},
  year    = {2024},
  doi     = {10.22331/q-2024-12-12-1563}
}

@article{Engel2019,
  author  = {Engel, Alexander and Smith, Graeme and Parker, Scott E.},
  title   = {Quantum algorithm for the {V}lasov equation},
  journal = {Physical Review A},
  volume  = {100},
  number  = {6},
  pages   = {062315},
  year    = {2019},
  doi     = {10.1103/PhysRevA.100.062315}
}

@article{Ameri2023,
  author  = {Ameri, Abtin and Ye, Erika and Cappellaro, Paola
             and Krovi, Hari and Loureiro, Nuno F.},
  title   = {Quantum algorithm for the linear {V}lasov equation with collisions},
  journal = {Physical Review A},
  volume  = {107},
  number  = {6},
  pages   = {062412},
  year    = {2023},
  doi     = {10.1103/PhysRevA.107.062412}
}

@article{Toyoizumi2024,
  author  = {Toyoizumi, Kiichiro and Yamamoto, Naoki and Hoshino, Kazuo},
  title   = {Hamiltonian simulation using the quantum singular-value
             transformation: Complexity analysis and application to the
             linearized {V}lasov--{P}oisson equation},
  journal = {Physical Review A},
  volume  = {109},
  number  = {1},
  pages   = {012430},
  year    = {2024},
  doi     = {10.1103/PhysRevA.109.012430}
}

@article{Miyamoto2024,
  author  = {Miyamoto, Koichi and Yamazaki, Soichiro and Uchida, Fumio
             and Fujisawa, Kotaro and Yoshida, Naoki},
  title   = {Quantum algorithm for the {V}lasov simulation of the
             large-scale structure formation with massive neutrinos},
  journal = {Physical Review Research},
  volume  = {6},
  number  = {1},
  pages   = {013200},
  year    = {2024},
  doi     = {10.1103/PhysRevResearch.6.013200}
}

@article{Higuchi2025,
  author  = {Higuchi, Hayato and Pedersen, Juan William and Toyoizumi, Kiichiro
             and Yoshikawa, Kohji and Kiumi, Chusei and Yoshikawa, Akimasa},
  title   = {Quantum calculation for two-stream instability and advection test
             of {V}lasov--{M}axwell equations: numerical evaluation of
             {H}amiltonian simulation},
  journal = {Journal of Plasma Physics},
  volume  = {91},
  number  = {4},
  pages   = {E116},
  year    = {2025},
  doi     = {10.1017/S0022377825100500}
}

@article{Einkemmer2024,
  author  = {Einkemmer, Lukas and Li, Qin and Wang, Li and Yang, Yunan},
  title   = {Suppressing instability in a {V}lasov--{P}oisson system by an
             external electric field through constrained optimization},
  journal = {Journal of Computational Physics},
  volume  = {498},
  pages   = {112662},
  year    = {2024},
  doi     = {10.1016/j.jcp.2023.112662}
}

@article{EinkemmerEtAl2025,
  author  = {Einkemmer, Lukas and Li, Qin and Mouhot, Cl{\'e}ment
             and Yue, Yukun},
  title   = {Control of instability in a {V}lasov--{P}oisson system
             through an external electric field},
  journal = {Journal of Computational Physics},
  volume  = {530},
  pages   = {113904},
  year    = {2025},
  doi     = {10.1016/j.jcp.2025.113904}
}

@article{VelozoRuiz2024,
  author  = {Velozo Ruiz, Anibal and Velozo Ruiz, Renato},
  title   = {Small Data Solutions for the {V}lasov--{P}oisson System
             with a Repulsive Potential},
  journal = {Communications in Mathematical Physics},
  volume  = {405},
  number  = {3},
  pages   = {80},
  year    = {2024},
  doi     = {10.1007/s00220-024-04970-3}
}

@article{Griffin-Pickering2021,
  author  = {Griffin-Pickering, Megan and Iacobelli, Mikaela},
  title   = {Global strong solutions in $\mathbb{R}^3$ for ionic
             {V}lasov--{P}oisson systems},
  journal = {Kinetic and Related Models},
  volume  = {14},
  number  = {4},
  pages   = {571--597},
  year    = {2021},
  doi     = {10.3934/krm.2021016}
}

@article{Strang1968,
  author  = {Strang, Gilbert},
  title   = {On the Construction and Comparison of Difference Schemes},
  journal = {SIAM Journal on Numerical Analysis},
  volume  = {5},
  number  = {3},
  pages   = {506--517},
  year    = {1968},
  doi     = {10.1137/0705041}
}

@article{Cheng1976,
  author  = {Cheng, C. Z. and Knorr, Georg},
  title   = {The integration of the {V}lasov equation in configuration space},
  journal = {Journal of Computational Physics},
  volume  = {22},
  number  = {3},
  pages   = {330--351},
  year    = {1976},
  doi     = {10.1016/0021-9991(76)90053-X}
}

@article{Jin2024PRL,
  author  = {Jin, Shi and Liu, Nana and Yu, Yue},
  title   = {Quantum Simulation of Partial Differential Equations via
             {S}chr{\"o}dingerization},
  journal = {Physical Review Letters},
  volume  = {133},
  number  = {23},
  pages   = {230602},
  year    = {2024},
  doi     = {10.1103/PhysRevLett.133.230602}
}

@misc{qiskit2024,
  author        = {Javadi-Abhari, Ali and Treinish, Matthew and Krsulich, Kevin
                   and Wood, Christopher J. and Lishman, Jake and Gacon, Julien
                   and Martiel, Simon and Nation, Paul D. and Bishop, Lev S.
                   and Cross, Andrew W. and Johnson, Blake R.
                   and Gambetta, Jay M.},
  title         = {Quantum computing with {Q}iskit},
  year          = {2024},
  eprint        = {2405.08810},
  archivePrefix = {arXiv},
  primaryClass  = {quant-ph},
  doi           = {10.48550/arXiv.2405.08810}
}

@article{Jin2025SIAM,
  author  = {Jin, Shi and Liu, Nana and Ma, Chuwen},
  title   = {{S}chr{\"o}dingerization based computationally stable algorithms
             for ill-posed problems in partial differential equations},
  journal = {SIAM Journal on Scientific Computing},
  volume  = {47},
  number  = {4},
  pages   = {B976--B1000},
  year    = {2025},
  doi     = {10.1137/24M1656116}
}

@article{Jin2026,
  author  = {Jin, Shi and Liu, Nana and Ma, Chuwen and Peng, Yizhe and Yu, Yue},
  title   = {On the {S}chr{\"o}dingerization method for linear non-unitary
             dynamics with optimal dependence on matrix queries},
  journal = {Communications in Mathematical Sciences},
  volume  = {24},
  number  = {2},
  pages   = {565--592},
  year    = {2026},
  doi     = {10.4310/cms.260103023937}
}

@article{Mottonen2005,
  author  = {M{\"o}tt{\"o}nen, Mikko and Vartiainen, Juha J.
             and Bergholm, Ville and Salomaa, Martti M.},
  title   = {Transformation of quantum states using uniformly controlled rotations},
  journal = {Quantum Information and Computation},
  volume  = {5},
  number  = {6},
  pages   = {467--473},
  year    = {2005},
  doi     = {10.26421/QIC5.6-5}
}

\clearpage
\appendix
\titleformat{\section}{\large\bfseries}{Appendix \thesection:}{0.5em}{}

\renewcommand{\thefigure}{A\arabic{figure}}
\renewcommand{\thetable}{A\arabic{table}}
\renewcommand{\theequation}{A\arabic{equation}}
\setcounter{figure}{0}
\setcounter{table}{0} 
\setcounter{equation}{0}

\section{Fourier-space initialisation and structured loading}
\label{app:fourier_space_loading}
In the Schr\"odingerisation framework, the auxiliary Fourier-space register encodes an initial profile in the warped phase variable \(p\). Its choice affects both the convergence approximation and the circuit complexity of the state preparation. Starting from the warped phase transformation \(\Phi(t,p)=e^{-p}f(t)\), the original equation is extended to \(p<0\) via initial condition \(\Phi(0,p)=e^{-|p|}f(0)\)~\cite{Jin2024PRL,Hu2024}. The natural choice of auxiliary variable \(f_p(p)=e^{-|p|}\) is continuous but not differentiable at \(p=0\) so the Fourier transform exhibit only algebraic decay in \(p\) (cf. Remark 5~\cite{Hu2024}).

A natural idea is to replace the cusp of \(e^{-|p|}\) by a smooth splice. For example, one may define~\cite{Jin2025SIAM}
\begin{equation}
f_p^{her}(p)=
\begin{cases}
e^{p}, & p<-1,\\[0.3em]
h(p), & -1\le p<0,\\[0.3em]
e^{-p}, & p\ge 0,
\end{cases}
\end{equation}
where \(h\) is a Hermite interpolation polynomial fitted at \(p=-1\) and \(p=0\). This  \(f_p^{her}(p)\) improves the regularity relative to \(e^{-|p|}\), but it remains only a near-optimal remedy.

\subsection{Error-function profile for optimal scaling}
\label{app:error_function}

To achieve optimal scaling, Jin \emph{et al.}~\cite{Jin2026} introduced an error-function profile
\begin{equation}
f_p^{\rm erf}(p)=e^{-p}\phi(p),
\qquad
\phi(p)=\frac{1+\operatorname{erf}(ap)}{2},
\label{eq:error_function}
\end{equation}
where \(\phi(p)\) interpolates smoothly between \(0\) on the negative side and \(1\) on the positive side. Unlike the original choice \(e^{-|p|}\), which has a cusp at \(p=0\), the error-function profile varies smoothly through the origin and is in fact analytic in \(p\). 

In practice, the sharpness parameter \(a\) is chosen from a prescribed tolerance \(\varepsilon\). Writing the switching factor as \(\phi(p)\), one requires \(|\phi(p)-1|\le \varepsilon\) for \(p\ge p^\ast\), which leads to
\begin{equation}
p^\ast = \frac{\sqrt{\log(1/\varepsilon)}}{a}.
\label{eq:p_star}
\end{equation}
A canonical choice of \(a = 2\sqrt{\log(1/\varepsilon)}\) gives \(p^\ast=1/2\). On the warped grid \(\{p_j\}\), the sampled error-function profile is then
\begin{equation}
f_{p,j}^{\rm erf} = e^{-p_j}\,\frac{1+\operatorname{erf}(a p_j)}{2}.
\label{eq:discrete_error_function}
\end{equation}
Unlike the original profile~\cite{Hu2024}, this vector is strongly asymmetric. In particular,
\begin{equation}
f_p^{\rm erf}(0)=\frac12,
\end{equation}
whereas \(f_p(0)=1\) for the original profile. For the error-function initialisation, the physical solution should be reconstructed from a positive \(p\)-slice lying beyond the transition region, where \(\phi(p)\approx 1\) and hence \(f_p^{\rm erf}(p)\approx e^{-p}\).

Once the error-function profile Eq.~\eqref{eq:error_function} is adopted, the next question is how to prepare the auxiliary \(p\)-register efficiently. A generic M\"ott\"onen state preparation operator~\cite{Mottonen2005}, such as the \texttt{initialize} instruction set~\cite{qiskit2024}, loads the normalised sampled vector exactly, but it treats the target as an arbitrary amplitude vector. 

Let
\begin{equation}
    N_p=2^{n_p}, \qquad \Delta p=\frac{2\pi R}{N_p}, \qquad 
    p_j=-\pi R+j\Delta p.
\end{equation}

\begin{remark}[Readout slice for the error-function profile.{\cite[Theorem~5.2]{Jin2026}}]
For \(e^{-|p|}\), \(p=0\) is a natural recovery point because \(f_p(0)=1\). For \(f_p^{\rm erf}\), however, \(p=0\) lies in the transition layer, where \(\phi(0)=1/2\). Recovery should instead be performed from a positive \(p\)-slice for which the switching factor has already saturated, so that
\[
    \phi(p_j)\approx 1,
    \qquad
    f_p^{\rm erf}(p_j)\approx e^{-p_j}.
\]

Using the grid notation defined above, we select a recovery index \(j_{\rm rec}\) such that \(p_{j_{\rm rec}}\) lies away from both the switching region and the right boundary. The admissible window is
\begin{equation}
    p_{\min}=p^\ast+\Lambda_+T+n_{\rm guard}\Delta p, \qquad
    p_{\max}=p_{N_p-1}-\Lambda_-T-n_{\rm guard}\Delta p,
\end{equation}
where \(p^\ast\) is the point beyond which \(|\phi(p)-1|\le \epsilon\). \(n_{\mathrm{guard}}\) is a non-negative integer specifying the number of auxiliary grid spacings reserved as a safety margin, while \(\Lambda_\pm T\) denote conservative bounds on the auxiliary-direction drift over the evolution time \(T\).

Here, the readout slice is selected by
\begin{equation}
    j_{\rm rec} = \operatorname*{arg\,min}_{0\le j<N_p}
    \left| p_j-\operatorname{clamp}(\pi,p_{\min},p_{\max})
    \right|.
\end{equation}
Note that the target \(p=\pi\) is not assumed to be universally optimal. At the selected slice, the full physical-grid vector
\begin{equation}
    \Phi(T,p_{j_{\rm rec}})
    \approx
    e^{-p_{j_{\rm rec}}} f(T),
\end{equation}
is obtained by fixing the auxiliary coordinate to \(p_{j_{\rm rec}}\).
\end{remark}

\begin{remark}[Nearly geometric structure of the positive tail]
The error-function profile Eq.~\eqref{eq:error_function}~\cite{Jin2026} has an additional discrete structure on the positive half of the \(p\)-grid. Since the positive half of the grid begins at \(j=N_p/2\),
\[
p_{N_p/2+k}=k\Delta p, \qquad k=0,\ldots,2^{n_p-1}-1 .
\]
Substituting \(j=N_p/2+k\) into Eq.~\eqref{eq:discrete_error_function} gives
\begin{equation}
f_{p,N_p/2+k}^{\rm erf} = e^{-k\Delta p} \frac{1+\operatorname{erf}(a k\Delta p)}{2}.
\end{equation}
The first factor is exactly geometric in the grid index \(k\),
\begin{equation}
e^{-k\Delta p} = \left(e^{-\Delta p}\right)^k = r^k, \qquad 
r:=e^{-\Delta p},
\end{equation}
while the second factor is a smooth switching function, which equals \(1/2\) at \(p=0\).  

For \(k\Delta p\gtrsim p^\ast\), however, the switching factor is close to one. Thus the positive tail satisfies
\begin{equation}
f_{p,N_p/2+k}^{\rm erf} \approx r^k .
\end{equation}
The auxiliary profile is not an arbitrary vector, but it consists of a nearly geometric positive tail, with a small non-geometric transition region near \(p=0\). 
\end{remark}

This offers possibilities for efficient state preparation, as follows.

\subsubsection{Geometric half-support loading}
\label{app:geometric_half_support_loading}

To exploit this structure, let \(m=n_p-1\) denote the number of qubits used to index the positive half of the \(p\)-register. Using the binary expansion
\begin{equation}
k=\sum_{l=0}^{m-1}2^l b_l,     \qquad      b_l\in\{0,1\},
\end{equation}
the geometric factor reads
\begin{equation}
r^k = \prod_{l=0}^{m-1} \left(r^{2^l}\right)^{b_l}.
\end{equation}

Define the normalised geometric tail state as a product state,
\begin{equation}
\ket{g_r} = \bigotimes_{l=0}^{m-1} \frac{\ket{0}+r^{2^l}\ket{1}}{\sqrt{1+r^{2^{l+1}}}}.
\end{equation}
This state can be prepared by applying one \(R_y\) rotation to each of the \(m\) lower qubits,
\begin{equation}
R_y(\theta_l), \qquad \theta_l = 2\tan^{-1}\!\left(r^{2^l}\right),
\qquad
l=0,\ldots,m-1.
\end{equation}
An additional \textsc{NOT} gate on the most significant \(p\)-qubit selects the nonnegative half of the grid. Thus the geometric half-support loader prepares
\begin{equation}
\ket{0}^{\otimes n_p} \longmapsto \ket{1}\ket{g_r},
\label{eq:geom_half_support}
\end{equation}
using only \(O(n_p)\) single-qubit rotations.

\subsubsection{Patched geometric loading}
\label{app:patched_geom_loading}

Let \(P=2^b\) be the number of positive grid points to patch, where \(b\le m\), and
\[
t_k := f_{p,N_p/2+k}^{\rm erf}, \qquad k = 0,\ldots,P-1.
\]
The normalised target patch, embedded in the \(m\)-qubit positive-half register, is
\begin{equation}
\lvert t_P^{(m)} \rangle := \frac{1}{\mathcal T_P} \sum_{k=0}^{P-1} t_k \lvert k \rangle_m,
\qquad
\mathcal T_P := \left(\sum_{k=0}^{P-1} |t_k|^2 \right)^{1/2},
\end{equation}
where \(\lvert k \rangle_m\) denotes the \(m\)-qubit computational basis state corresponding to the integer \(k\).

On this selected subspace, the circuit first unprepares the local geometric state on the first \(b\) qubits and then prepares the exact target patch state \(\lvert t_P^{(m)}\rangle\).
\begin{equation}
\widetilde g_k
=
\begin{cases}
\displaystyle
B_P
\frac{t_k}
{\left(\sum_{l=0}^{P-1}|t_l|^2\right)^{1/2}},
& 0\le k<P,\\[1.0em]
g_k, & k\ge P,
\end{cases}
\end{equation}
where the geometric probability assigned to the first \(P\) amplitudes is
\begin{equation}
B_P^2 = \sum_{k=0}^{P-1}|g_k|^2 .
\end{equation}
The resulting loader is a product-state geometric loader plus a local exact correction on \(P\) amplitudes.

\paragraph{Circuit description}


Let \(G_b\) denote the geometric product loader restricted to the first
\(b\) positive-half qubits and let
\begin{equation}
T_b\lvert 0\rangle^{\otimes b} = \frac{1}{\mathcal T_P} \sum_{k=0}^{P-1} t_k \lvert k\rangle.
\end{equation}
After the full geometric loader has prepared \(\lvert 1\rangle\lvert g_r\rangle\), the circuit applies the zero-controlled operation \(\mathcal C_0\!\left(T_bG_b^\dagger\right)\), where the controls are \(q_b,\ldots,q_{m-1}\). Thus, the local geometric state is unprepared and replaced by the
exact patch state:
\begin{equation}
\lvert g_r^{(b)}\rangle \;\xrightarrow{\;G_b^\dagger\;}\;
\lvert 0\rangle^{\otimes b} \;\xrightarrow{\;T_b\;}\;
\frac{1}{\mathcal T_P} \sum_{k=0}^{P-1} t_k\lvert k\rangle,
\end{equation}
with other positive-half blocks unchanged. Hence, the patched loader prepares
\begin{equation}
\lvert 0\rangle^{\otimes n_p} \longmapsto \lvert 1\rangle\lvert g_r\rangle
\longmapsto \lvert 1\rangle\lvert \widetilde g\rangle ,
\end{equation}
where the second map is the zero-controlled \(\mathcal C_0\!\left(T_bG_b^\dagger\right)\). 

\subsection{Benchmark and scaling}
\label{app:benchmark_scaling}

\begin{figure}[htbp]
\centering
\includegraphics[width=0.6\linewidth]{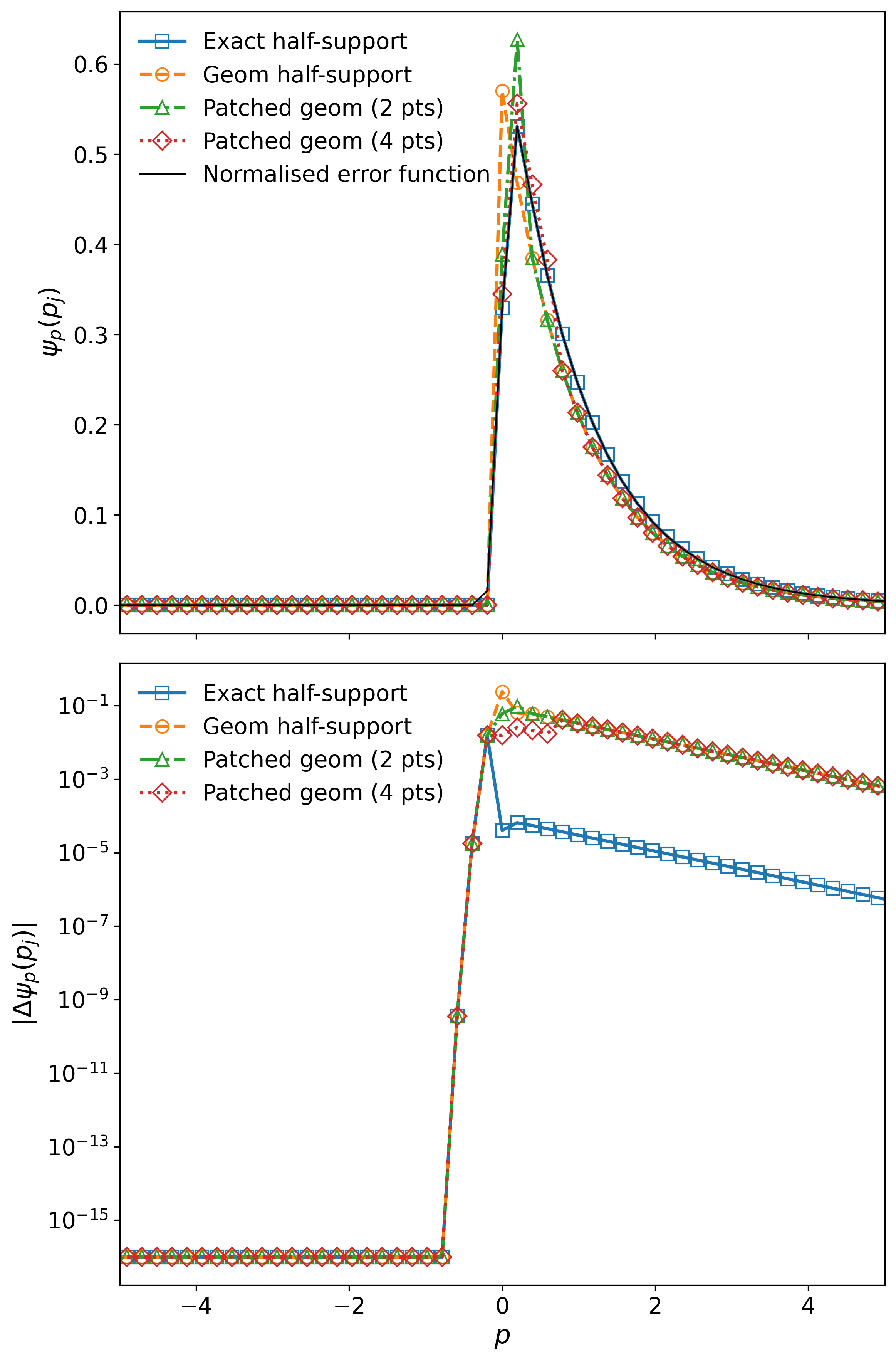}
\caption{Amplitude \(\psi_p(p_j)\) and pointwise-error \(|\Delta\psi_p(p_j)|\) plotted against \(p\in(-5,5)\) for structured \(p\)-register loaders with \(n_p=7\) and \(R=4\).}
\label{fig:patched_loader_comparison}
\end{figure}

We quantify the loader error using the state fidelity
\begin{equation}
\mathcal F(\psi,\varphi) := |\langle \psi \,|\, \varphi\rangle|^2,
\end{equation}
where \(\lvert \varphi\rangle\) is the normalised sampled error-function target and \(\lvert \psi\rangle\) is the state produced by the loader. Fig.~\ref{fig:patched_loader_comparison} shows a comparison of amplitude \(\psi_p(p_j)\) and pointwise-error \(|\Delta\psi_p(p_j)|\) for structured \(p\)-register loaders for \(n_p=7\) and \(R=4\). The geometric half-support loader captures the positive exponential tail but has localised error near \(p=0\). Patched geometric loaders reduce this transition-region error by applying an exact local correction to the first \(P\) positive-grid amplitudes.

Table~\ref{tab:p_loader_scaling} compares two representative grids with the same \(R=4\). For \(n_p=5\), the grid spacing is \(\Delta p =\pi/4,\) and a four-point patch \(P=4\) gives fidelity \(\mathcal F \approx 0.9994\). For \(n_p=7\), the grid is finer \(\Delta p =\pi/16,\) and \(P=4\) gives fidelity \(\mathcal F \approx 0.9931\). To refine the \(p\)-grid, one may increase \(P\) to cover more of the switching region.

\begin{table}[htbp]
\centering
\caption{Loader fidelities \(\mathcal F\) and circuit metrics for \(R=4\) and \(\varepsilon=10^{-6}\). Fidelity values are rounded to four decimal places. The exact tree loaders grow rapidly with \(n_p\). The patched loaders trade a local exact correction near \(p=0\) for improved fidelity.}
\label{tab:p_loader_scaling}
\small
\setlength{\tabcolsep}{3.5pt}
\begin{tabular}{|l|cccc|cccc|}
\hline
& \multicolumn{4}{c|}{\(n_p=5\)} 
& \multicolumn{4}{c|}{\(n_p=7\)} \\
\cline{2-5}\cline{6-9}
Loader 
& \(\mathcal F\) & Depth & Size & CNOT
& \(\mathcal F\) & Depth & Size & CNOT \\
\hline

Exact tree 
& 1.0000 & 548 & 817 & 290
& 1.0000 & 4073 & 7055 & 2872 \\

Exact half-support tree 
& 1.0000 & 408 & 576 & 196
& 0.9998 & 3086 & 4896 & 1860 \\

Geometric half-support 
& 0.8986 & 1 & 5 & 0
& 0.9282 & 1 & 7 & 0 \\

Patched geometric, \(P=2\) 
& 0.9850 & 85 & 127 & 40
& 0.9761 & 132 & 219 & 80 \\

Patched geometric, \(P=4\) 
& 0.9994 & 233 & 328 & 124
& 0.9931 & 1016 & 1525 & 576 \\
\hline
\end{tabular}

\end{table}

A generic exact tree on \(N_p = 2^{n_p}\) amplitudes requires \(O(2^{n_p})\) controlled rotations, and the circuit metrics grow rapidly as \(n_p\) increases from \(5\) to \(7\). By contrast, the geometric half-support loader uses only \(m = n_p - 1\) independent one-qubit rotations together with a single \textsc{NOT} gate, which represents a clear scaling advantage, but at the cost of fidelity.

The patched loader supplements this geometric construction with a local exact correction acting on the first \(P\) positive amplitudes. To ensure that the patch spans the entire switching region, one should choose \(P\) such that \(P\Delta p \gtrsim p^\ast\), or equivalently,
\begin{equation}
P \gtrsim \frac{p^\ast 2^{n_p}}{2\pi R}.
\end{equation}

\begin{figure}[htbp]
\centering
\includegraphics[width=0.95\linewidth]{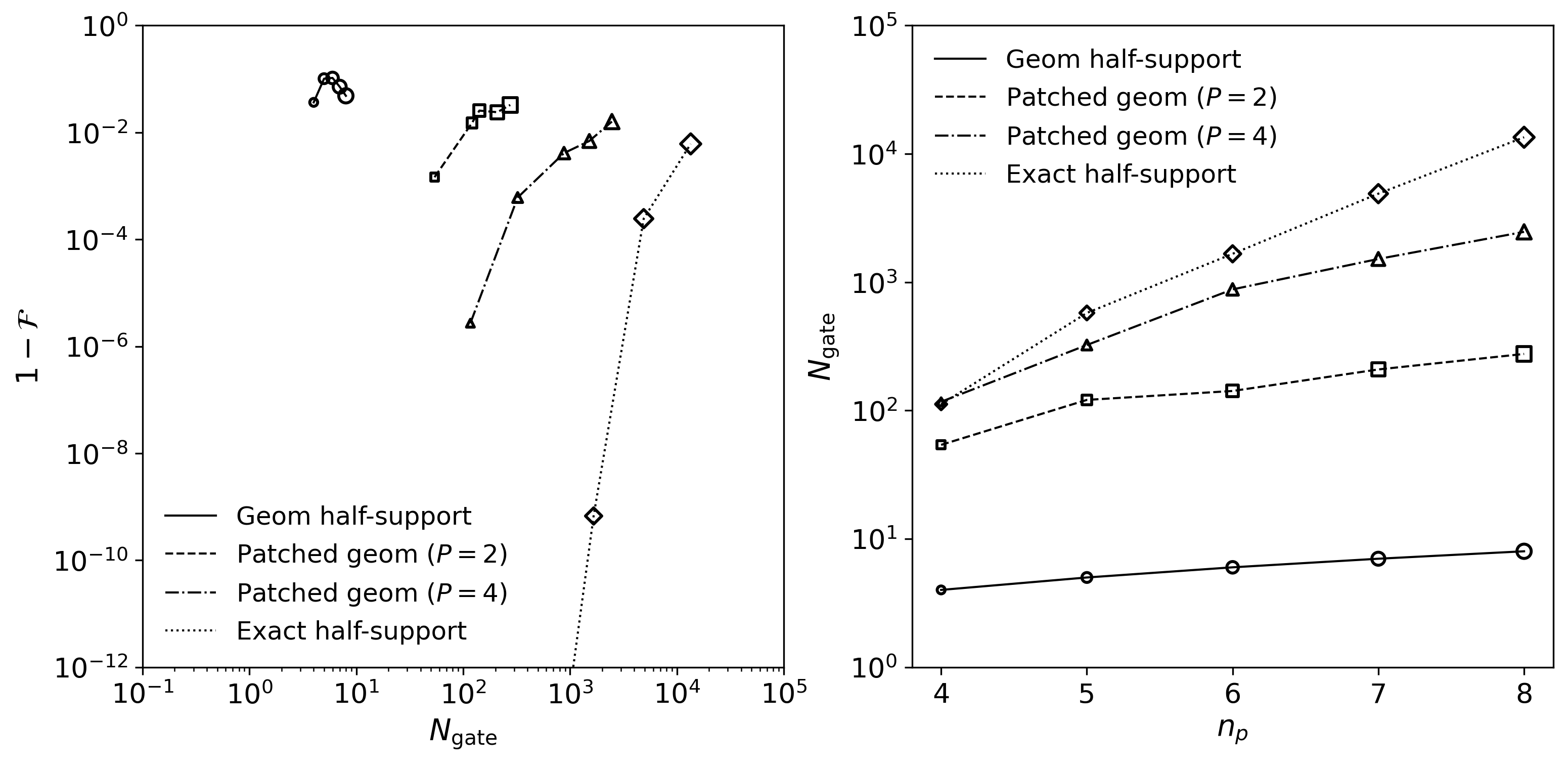}
\caption{Accuracy--cost and cost-scaling comparison for the structured \(p\)-register loaders as \(n_p\) varies from \(4\) to \(8\), with larger markers indicating larger \(n_p\). (Left) Error \(1-\mathcal F\) versus decomposed gate count \(N_{\rm gate}\), with each curve corresponding to a fixed loader. (Right) Decomposed circuit size as a function of \(n_p\) for the same loaders.}
\label{fig:p_loader_accuracy_cost_scaling}
\end{figure}

The geometric part of the construction is scalable, while the patch introduces a controlled local state-preparation overhead whose cost depends on both \(P\) and the number of control qubits. Fig.~\ref{fig:p_loader_accuracy_cost_scaling} summarises this accuracy--cost tradeoff. The left panel plots the error \(1-\mathcal F\) against the decomposed gate count \(N_{\rm gate}\), which is the total number of elementary gates after circuit decomposition into Qiskit's standard one- and two-qubit gate set. The right panel plots the same circuit-size metric as a function of \(n_p\). The geometric loader remains small because it uses only one-qubit rotations, while the patched loaders improve fidelity at the price of controlled-patch overhead.

\subsection{Reality of finite-register readout}
\label{app:erf_readout_reality_check}

The preceding discussion shows that the error-function profile has two attractive features in that it removes the cusp of \(e^{-|p|}\), and its positive tail is nearly geometric on the discrete \(p\)-grid. These properties make it appealing for structured state preparation; in particular, the geometric half-support achieves fidelity up to \(\mathcal{F} \approx0.93\) using \(\mathcal{O}(n_p)\) single-qubit rotations arranged in \(\mathcal{O}(1)\) circuit depth. However, its finite-register readout has additional practical considerations.

For the original profile \(e^{-|p|}\), the central slice \(p=0\) is a natural readout point because \(f_p(0)=1\). By contrast, the error-function profile is one-sided, and \(p=0\) lies inside the switching layer where \(\phi(0)=1/2\). Thus the error-function profile requires readout from a positive \(p\)-slice beyond the switching region.

Table~\ref{tab:erf_reality_check} reports a reduced proof-of-concept comparison for the centred-grid double-well Vlasov test. All runs use
\[
    n_x=n_v=2,\qquad R=4,\qquad N_t=5,\qquad \Delta t=0.01.
\]
The original profile is read out at \(p=0\), while the error-function profile uses the bounded-\(\pi\) rule.

\begin{table}[htbp]
\centering
\caption{Finite-register comparison of the original auxiliary profile and the error-function profile.}
\label{tab:erf_reality_check}
\begin{tabular}{ccccc}
\toprule
\(n_p\)
& original rel. \(L^2\)
& original mass
& erf rel. \(L^2\)
& erf mass \\
\midrule
5 & \(6.12\times 10^{-3}\) & 1.0154 & \(4.56\times 10^{-3}\) & 1.0140\\
6 & \(9.02\times 10^{-3}\) & 1.0142 & \(1.18\times 10^{-2}\) & 1.0063\\
7 & \(1.22\times 10^{-2}\) & 1.0122 & \(1.72\times 10^{-2}\) & 1.0019\\
\bottomrule
\end{tabular}
\end{table}

The \(n_p=5\) case shows that the error-function profile can perform well with the \(p=\pi\) slice. However, increasing \(n_p\) does not produce a monotone improvement in this reduced test. Therefore \(p=\pi\) is not a universally optimal readout slice, and it is unclear whether a universally optimal slice exists.

The ambiguity of selecting the readout slice within the admissible interval \((p_{\min},p_{\max})\) introduces an additional
practical reality in the direct application of the error-function profile since the number of discrete grid points in this interval may scale as
\(\mathcal{O}(2^{n_p})\). Hence, the original profile \(e^{-|p|}\), or even an interpolated polynomial profile, could be more practical for PDE solution simply because its recovery slice is fixed a priori at \(p=0\). 

\section{Gate-level details for Block 2 arithmetic}
\label{app:block2_arithmetic}

\renewcommand{\thefigure}{B\arabic{figure}}
\renewcommand{\thetable}{B\arabic{table}}
\renewcommand{\theequation}{B\arabic{equation}}
\setcounter{figure}{0}
\setcounter{table}{0} 
\setcounter{equation}{0} 

This appendix expands the coherent field-register construction used in Block~2. The goal is to implement the reversible map
\begin{equation}
\lvert j\rangle_X\lvert 0\rangle_E \longmapsto \lvert j\rangle_X\lvert E_{\rm fp}(j)\rangle_E,
\qquad
E_{\rm fp}(j)=\alpha_{\rm fp}c(j)+\beta_{\rm fp}c(j)^3,
\end{equation}
where \(c(j)\) is the centred signed integer associated with the encoded position index \(j\). All arithmetic is reversible and all temporary workspaces are uncomputed unless explicitly stated otherwise.

\paragraph{Step 1: Preparation of the centred signed register.}
The position index is first copied into an \(n_x\)-qubit work register \(c\),
\begin{equation}
\lvert j\rangle_X\lvert 0\rangle_c
\longmapsto
\lvert j\rangle_X\lvert j\rangle_c .
\end{equation}
Then we flip the most-significant bit of \(c\) and interpret the bit string in two's-complement form, giving the centred signed integer
\begin{equation}
c(j)=j-2^{n_x-1}.
\end{equation}
This way, the unsigned range \(j=0,\ldots,2^{n_x}-1\) is mapped to the centred
signed range
\[
-2^{n_x-1},\ldots,2^{n_x-1}-1.
\]

\paragraph{Step 2: Magnitude extraction.}
The bit string stored in \(c\) is copied into an \(n_x\)-qubit unsigned magnitude register \(c_{\rm abs}\). Let \(s\) denote the sign bit of \(c\). We map
\begin{equation}
\lvert c\rangle_c\lvert 0\rangle_{c_{\rm abs}} \longmapsto
\lvert c\rangle_c\lvert |c|\rangle_{c_{\rm abs}},
\end{equation}
where the target register is left unchanged when \(s=0\), and is two's-complement negated when \(s=1\). Thus \(c_{\rm abs}\) stores the unsigned magnitude of \(c\), while the signed source register \(c\) is preserved for later use.

\paragraph{Step 3: Square and cubic workspaces.}
Using reversible multiplication, the magnitude register is promoted first to
a square workspace
\begin{equation}
\ket{|c|}_{c_{\mathrm{abs}}}\ket{0}_{\mathrm{sq}} \longmapsto
\ket{|c|}_{c_{\mathrm{abs}}}\ket{|c|^2}_{\mathrm{sq}},
\label{eq:block2_step3_square}
\end{equation}
and then to a cubic workspace
\begin{equation}
\ket{|c|}_{c_{\mathrm{abs}}}\ket{|c|^2}_{\mathrm{sq}}\ket{0}_{\mathrm{cube}} \longmapsto \ket{|c|}_{c_{\mathrm{abs}}}\ket{|c|^2}_{\mathrm{sq}}\ket{|c|^3}_{\mathrm{cube}}.
\label{eq:block2_step3_cube}
\end{equation}

The intermediate square register is then uncomputed. Although the
implementation may allocate a larger register \({\rm cube}\), only the
low-order \(3n_x\)-qubit subregister is needed to store the signed cubic
value. We denote this subregister by
\begin{equation}
C := {\rm cube}_{[0:3n_x-1]} .
\end{equation}

\paragraph{Step 4: Sign restoration of the cubic term.}
The register \(\mathrm{cube}\) now contains the unsigned magnitude \(|c|^3\). To obtain the signed cubic value \(c^3\), the circuit applies a
controlled two's-complement negation, controlled on the sign bit \(s\), such that
\begin{equation}
\lvert s\rangle_{\mathrm{MSB}(c)}\lvert y\rangle_C
\longmapsto
\begin{cases}
\lvert 0\rangle_{\mathrm{MSB}(c)} \lvert y\rangle_C, & s=0,\\[0.4em]
\lvert 1\rangle_{\mathrm{MSB}(c)} \lvert -y\rangle_C, & s=1.
\end{cases}
\end{equation}
Thus,
\begin{equation}
\lvert c\rangle_c\lvert |c|^3\rangle_C \longmapsto 
\lvert c\rangle_c \lvert c^3\rangle_C .
\end{equation}
At the gate level, the controlled negation is implemented by a sign-controlled
layer of \textsc{NOT} gates on \(C\), followed by a sign-controlled increment
by one, with the carry ancillas uncomputed at the end. The higher-order
padding bits of the full \(\mathrm{cube}\) workspace are left unchanged.

\paragraph{Step 5: Weighted accumulation into the field register.}
The output register \(E\) is assembled by accumulating the signed linear and
cubic contributions into a clean \(n_E\)-qubit register,
\begin{equation}
\lvert c\rangle_c \lvert c^3\rangle_C \lvert 0\rangle_E \longmapsto
\lvert c\rangle_c \lvert c^3\rangle_C \lvert \alpha_{\rm fp}c+\beta_{\rm fp}c^3\rangle_E .
\end{equation}
This accumulation is implemented bit by bit using the two's-complement
expansions of the signed source registers. For the linear term,
\begin{equation}
c = \sum_{i=0}^{n_x-2}2^i c_i - 2^{n_x-1}c_{\rm msb},
\end{equation}
and so
\begin{equation}
\alpha_{\rm fp}c = \sum_{i=0}^{n_x-2} c_i\left(\alpha_{\rm fp}2^i\right) +
c_{\rm msb}\left(-\alpha_{\rm fp}2^{n_x-1}\right).
\end{equation}
Thus, for each non-sign bit \(c_i\), the circuit conditionally adds the shifted constant \(\alpha_{\rm fp}2^i\) into \(E\), while for the sign bit it conditionally adds the two's-complement constant \(-\alpha_{\rm fp}2^{n_x-1}\). This uses \(n_x\) controlled constant-addition blocks.

The cubic term is accumulated in the same way from the signed \(3n_x\)-qubit register
\begin{equation}
C = \sum_{i=0}^{3n_x-2}2^i C_i - 2^{3n_x-1}C_{\rm msb},
\end{equation}
and
\begin{equation}
\beta_{\rm fp}C = \sum_{i=0}^{3n_x-2} C_i\left(\beta_{\rm fp}2^i\right) +
C_{\rm msb}\left(-\beta_{\rm fp}2^{3n_x-1}\right).
\end{equation}
Therefore, the cubic term requires \(3n_x\) controlled constant-addition blocks. In total, Step~5 uses \(n_x+3n_x=4n_x\) controlled constant-addition blocks.

Note that each controlled constant-addition block has the same structure. First, the appropriate shifted classical constant is loaded into the \(n_E\)-qubit workspace
\begin{equation}
W=(\mathrm{tmp},\mathrm{sq},w_{\mathrm{msb}},\mathrm{wpad}).
\end{equation}
Next, a CDKM ripple-carry adder adds this workspace register into \(E\), controlled by the relevant source bit. Finally, the workspace preparation and carry initialisation are uncomputed, so that only the updated output register \(E\) remains.

\section{Sensitivity to spatial grid}
\label{app:grid_sensitivity}
\renewcommand{\thefigure}{C\arabic{figure}}
\setcounter{figure}{0}

The original 1D1V implementation samples the initial phase-space distribution on a cell-centred spatial grid (Eq.~\eqref{eq:x_v_grids}), whereas the fixed-point acceleration field is evaluated using centred-integer coordinates corresponding to a nodal grid. To assess the numerical sensitivity of the circuit validation to this half-cell offset, we performed an additional reduced 1D1V study.

The test uses the following setting:
\[
n_x=n_v=3,\qquad n_p=5,\qquad R=4,\qquad
N_t=1,\qquad \Delta t=0.05,
\]
with \(L_x=L_v=6\), fixed-point precision \(s=6\), and zero mean drift \(v_0=0\). For this configuration, \(\alpha_{\mathrm{fp}}=48, \beta_{\mathrm{fp}}=-27\).

Two cases were considered:
\begin{enumerate}
\item Cell-centred convention, in which the initial distribution is sampled on a cell-centred spatial grid while the field is evaluated on the nodal grid.
\item Nodal convention, in which both the initial distribution and the field use nodal spatial coordinates.
\end{enumerate}

\begin{figure}[htbp]
\centering
\includegraphics[width=0.95\linewidth]{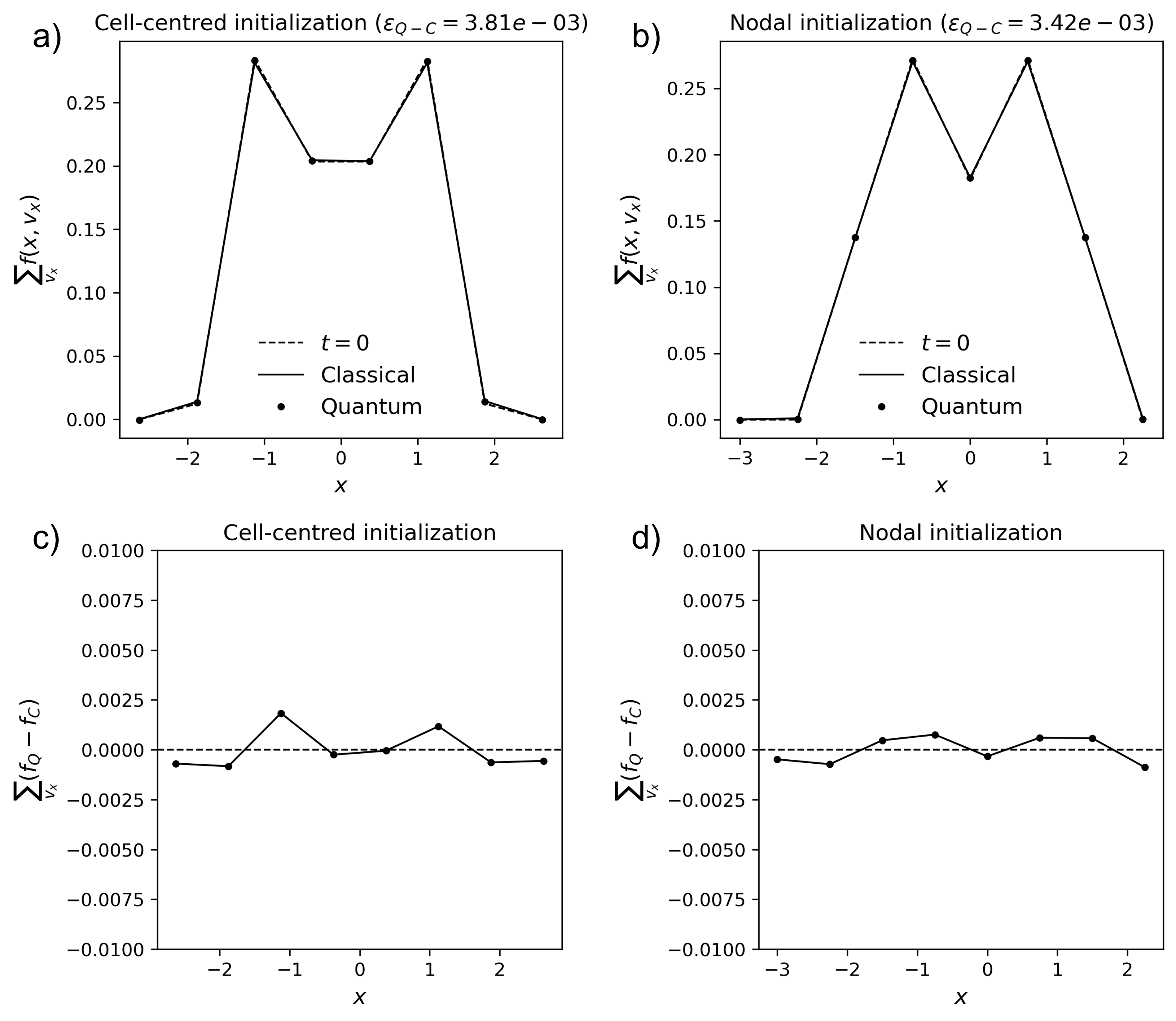}
\caption{
Sensitivity of the reduced 1D1V validation test to the
spatial grid convention for \(T=0.05\).
(a) Spatial marginal for the original implementation, in which the initial distribution is sampled on a cell-centred spatial grid.
(b) Spatial marginal for the consistent nodal grid initialization.
(c) Quantum--classical difference in the spatial marginal for the cell-centred initialization.
(d) Quantum--classical difference in the spatial marginal for the nodal initialization.
}
\label{fig:grid_sensitivity}
\end{figure}

Figure~\ref{fig:grid_sensitivity} compares the spatial
marginals for the two cases. In both conventions, the
recovered quantum marginal closely follows the
scheme-matched classical reference. The reported
quantum--classical discrepancies are \(\epsilon_{Q-C}=3.81\times10^{-3}\) for the original cell-centred initialization and \(\epsilon_{Q-C}=3.42\times10^{-3}\) for the nodal initialization. The corresponding Aer MPS execution times are \(9.29\) and \(9.69\) minutes, respectively, indicating that the classical emulation cost is also broadly insensitive to the choice of initialization grid in this reduced test.

The two grid conventions produce different discretised distributions, with an approximately \(17\%\) relative \(L_2\) difference between the classical solutions after periodic spatial-grid alignment. A comparable difference is already present in the initial distributions, reflecting the sensitivity of this coarse grid to spatial sampling. Nevertheless, both conventions yield similar quantum--classical discrepancies and MPS execution times. Thus, the grid convention affects the discretised solution but has limited impact on circuit validation and MPS emulation cost in this reduced test.

\end{document}